\documentclass[journal]{IEEEtran}
\usepackage{amsmath}

\usepackage{amssymb}
\usepackage{bm}
\usepackage{color}
\usepackage{graphicx}
\usepackage{empheq}
\usepackage[linesnumbered,ruled,lined]{algorithm2e}
\usepackage[outdir=./]{epstopdf}
\usepackage{caption}
\usepackage{lipsum}
\usepackage{cite}
\usepackage{multirow}
\usepackage{subcaption}
\usepackage{fancyhdr}
\usepackage{amsmath, amsthm, amssymb}
\SetKwInput{Input}{Input}
\SetKwInput{Output}{Output}

\usepackage{array}
\newcolumntype{L}[1]{>{\raggedright\let\newline\\\arraybackslash\hspace{0pt}}m{#1}}
\newcolumntype{C}[1]{>{\centering\let\newline\\\arraybackslash\hspace{0pt}}m{#1}}
\newcolumntype{R}[1]{>{\raggedleft\let\newline\\\arraybackslash\hspace{0pt}}m{#1}}
\IEEEoverridecommandlockouts
\newtheorem{prop}{Proposition}

\begin{document}

	\title{Max-min Fairness of $K$-user Cooperative
Rate-Splitting in MISO Broadcast Channel
with User Relaying}

				\author{
			
			\IEEEauthorblockN{
				Yijie~Mao, \IEEEmembership{Member,~IEEE,}   
				Bruno~Clerckx, \IEEEmembership{Senior Member,~IEEE,} 
				Jian~Zhang, 
				Victor~O.K.~Li, \IEEEmembership{Life Fellow,~IEEE,} 
				and~Mohammed~Arafah, \IEEEmembership{Member,~IEEE}  
	     \thanks{Manuscript received October 17, 2019; revised April 18, 2020; accepted June 10, 2020.}
		 \thanks{Y. Mao and B. Clerckx are with Imperial College London, London SW7 2AZ, UK (email: y.mao16@imperial.ac.uk; b.clerckx@imperial.ac.uk).}		 	
		 \thanks{J. Zhang is with Xidian University, Xi'an, China  (email: j.zhang18@imperial.ac.uk).}
		 \thanks{V.O.K. Li is with The University of Hong Kong, Hong Kong, China (email: vli@eee.hku.hk).}
		 \thanks{M. Arafah is with King Saud University, Riyadh, Saudi Arabia (email: arafah@ksu.edu.sa).}
	}}

\maketitle

\thispagestyle{empty}
\pagestyle{empty}
\begin{abstract}
Cooperative Rate-Splitting (CRS) strategy,  relying on linearly precoded rate-splitting at the transmitter and opportunistic transmission of the common message by the relaying user,  has recently been shown to outperform typical Non-cooperative Rate-Splitting (NRS), Cooperative Non-Orthogonal Multiple Access (C-NOMA) and Space Division Multiple Access (SDMA) in a two-user Multiple Input Single Output (MISO) Broadcast Channel (BC) with user relaying. In this work, the existing two-user CRS transmission strategy is generalized to the  $K$-user case. We study the problem of jointly optimizing the precoders, message split, time slot allocation, and relaying user scheduling with the objective of  maximizing the minimum rate among users subject to a  transmit power constraint at the base station. 
As the  user scheduling problem is discrete  and the entire problem is non-convex, we  propose a two-stage low-complexity algorithm to solve the problem.   Both centralized and decentralized  relaying protocols based on selecting $K_1$ ($K_1<K$) strongest users are first proposed followed by a Successive Convex Approximation (SCA)-based algorithm  to  jointly optimize the time slot, precoders and message split.  Numerical results show that by applying the proposed two-stage algorithm, the worst-case achievable rate achieved by CRS is significantly increased over that of NRS and SDMA in a wide range of  network loads (underloaded and overloaded regimes) and user deployments (with a diversity of channel strengths).  Importantly, the proposed SCA-based algorithm dramatically reduces the computational complexity without any rate loss compared with the  conventional algorithm in the literature of CRS.  Therefore, we conclude that the proposed $K$-user CRS combined with the two-stage algorithm is more powerful  than the existing  transmission schemes. 
\end{abstract}

\section{Introduction}
\IEEEPARstart{D}{riven} by the exponential rise in the volume of wireless traffic and increasing demands for higher data rates, considerable attention has been drawn to the design of innovative solutions   for future wireless communication systems. Among numerous candidate technologies, Rate-Splitting Multiple Access (RSMA), a novel multiple access scheme proposed in the recent work \cite{mao2017rate}, has been shown to be a promising and powerful strategy  to boost the lower layers of next-generation communication systems.
 By using  linearly precoded Rate-Splitting (RS) at the transmitter to split the user messages into common and private parts, and using Successive Interference Cancellation (SIC) at all receivers to sequentially decode the encoded common streams and the private streams, RSMA  manages interference in a more versatile and robust manner than Space Division Multiple Access (SDMA) that fully treats residual interference as noise and power-domain  Non-Orthogonal Multiple Access (NOMA) that fully decodes interference.  Therefore, RSMA unifies and outperforms SDMA and NOMA from both spectral and energy efficiency perspectives \cite{mao2019TCOM, mao2018EE,mao2018networkmimo,Jian2019CRS,bruno2019wcl}. 
 
 The idea of RS was first introduced in a two-user Single-Input Single-Output (SISO) Interference Channel (IC) \cite{TeHan1981} and was further developed in recent works \cite{RS2015bruno,RSintro16bruno,RS2016hamdi,RS2016joudeh,enrico2016bruno,enrico2017bruno,chenxi2017brunotopology,Medra2018SPAWC,Lu2018MMSERS,bruno2019wcl} for modern Multiple Input Multiple Output (MIMO) Broadcast Channels (BCs). Among those works,  1-layer RS, the simplest building block of the framework of RSMA,  has been studied and shown to be beneficial in the multi-antenna setup\footnote{ In the rest of the paper, ``1-layer RS" will be referred to by ``RS" for simplicity. }. RS splits the message of each user into one common and one private part. The common parts of all users are combined and encoded into one common stream to be decoded by all users while the privates parts are independently encoded and decoded by the corresponding users. RS superimposes the common stream on top of the private streams, and broadcasts to the users.
 In contrast to NOMA,  RS does not require any user ordering or grouping at the transmitter and only one single layer of SIC is required at each user to decode and remove the common stream before decoding the intended private stream.  The literature of RS in multi-antenna BC starts from an information theoretic perspective where  RS is shown to achieve the optimal sum Degree of Freedom (DoF) \cite{RS2016hamdi} followed by the entire DoF region \cite{enrico2017bruno} of the $K$-user underloaded Multiple Input Multiple Out (MISO) BC with imperfect Channel State Information at the Transmitter (CSIT). Motivated by the DoF results showing the multiplexing
gains of RS at high Signal-to-Noise Ratio (SNR), a number of recent works study the performance of RS in
the finite SNR regime of MISO BC with both perfect CSIT \cite{mao2017rate,SYang2018SPAWC,mao2018EE} and imperfect CSIT \cite{ RS2016hamdi,RSintro16bruno,RS2016joudeh,chenxi2017brunotopology,Medra2018SPAWC,Lu2018MMSERS}. Besides the typical MISO BC, the benefits of RS has been further exploited  in the finite SNR regime of  massive MIMO \cite{Minbo2016MassiveMIMO, AP2017bruno}, millimeter wave systems \cite{minbo2017mmWave,Kola2018SPAWC}, multigroup multicasting \cite{hamdi2015multicasting, Tervo2018SPAWC}, multi-cell Coordinated Multipoint Joint Transmission (CoMP) \cite{mao2018networkmimo}, Simultaneous Wireless Information and Power Transfer (SWIPT) \cite{mao2019swipt},   joint unicast and multicast transmission \cite{mao2019TCOM}, Cloud Radio Access Network (C-RAN) \cite{Ahmad2018SPAWC} as well as mutli-pair Decode-and-Forward (DF) full-duplex relay channel \cite{MultiPairRS2018Tassos}. 
As the common stream in RS is required to be decoded by all users, its achievable rate is limited by the
rate of the worst-case user. If the users experience heterogeneous channel strengths from BS, the rate of  the common stream (which is also known as common rate) may drop.  To overcome the limitation, one promising solution is to combine RS with cooperative transmission using user relaying.

Cooperative user relaying has been identified as one promising solution for effectively combating the shadowing effects to extend the radio coverage and  improve the channel capacity simultaneously \cite{relay2015Zhang}. By appointing one or more relaying users to assist in receiving data from BS and  forwarding it to other users, the spatial diversity is boosted without deploying dedicated relaying stations. BC has been shown to be well suited for cooperative user relaying in the literature of Relay BC (RBC)  \cite{SISOBC2006TIT,SISORBC2007TIT1,SISORBC2007TIT2} from an information theoretical perspective due to its capability to simultaneously serve  relaying users and other users in the same resource blocks. By incorporating superposition coding at the transmitter and Decoding-and-Forward (DF) at relaying users, RBC communication networks where  the transmitter sends information to a number of users that cooperate by exchanging information, significantly improve the rate region of typical SISO BC achieved by Dirty Paper Coding (DPC) or superposition coding \cite{SISORBC2007TIT2}. The benefits of RBC have been further explored in \cite{multiAnCRS2011Globecom,CRSMIMO2011TCOM } considering  multiple transmit  antennas at  BS in order to embrace the  spatial multiplexing gain of multi-antenna BC over single-antenna BC. In \cite{multiAnCRS2011Globecom}, the achievable rate region of two-user  multi-antenna RBC with Zero-Forcing (ZF)-DPC is shown to be larger than that of ZF-DPC in the two-user MIMO BC without user relaying. A low-complex ZF beamforming with closed-form optimal power allocation is proposed in \cite{CRSMIMO2011TCOM} for MIMO RBC.


Motivated by the recent findings about the performance benefits of RS over SDMA in multi-antenna BC, it is worth to explore the benefits of  incorporating  RS in multi-antenna RBC, which has only recently emerged.  Indeed, \cite{Jian2019CRS} recently proposed a linearly precoded two-user Cooperative RS (CRS) transmission framework by allowing one of the two users to opportunistically forward its decoded common message to the other user. The ``opportunistic" transmission comes from  dynamic time slot allocation for the two transmission phases (namely, direct transmission phase from BS to the two users  and cooperative transmission phase from the relaying user to the other user). It has been demonstrated in \cite{Jian2019CRS} that the proposed two-user CRS is more general than typical RS with only direct transmission, Cooperative-NOMA (C-NOMA),  cooperative RS with equal time slot allocation as well as  SDMA, and outperforms them all.

\subsection{Motivations and Contributions}
The objective of this work is to generalize the two-user CRS framework proposed in \cite{Jian2019CRS}  to a $K$-user setup. This brings a number of issues to overcome:
\begin{itemize}
	\item In the algorithm proposed in \cite{Jian2019CRS},  the optimization of time slot allocation for the two transmission phases is separated from the optimization of  precoders and message split. The  time slot allocation is achieved via one dimensional exhaustive search while the precoder and the message split are jointly optimized for each given time slot allocation scheme. The computational complexity of the algorithm is relatively high.  A first question arising is therefore ``Is there any simpler algorithm that derives the optimal time slot allocation without requiring an exhaustive search?"
	
	\item In the direct transmission phase of CRS, the encoded common stream and the private streams for all users are superimposed and broadcast to the users simultaneously. Each user decodes the common stream and the intended private stream sequentially with the assistance of SIC. 	As all users first decode the common stream, all of them are potential candidates for forwarding the common stream in the cooperative transmission phase. One can therefore wonder how to smartly select one or more relaying users?  If the relaying users are not properly selected, the cooperative transmission phase can be useless. For example,  if the user with the worst transmission rate is selected as a relaying user, the cooperative transmission from the worst-case user to other users cannot improve the achievable common rate   at all due to the fact that the achievable rate of the common stream  is limited by the  worst-case achievable rate. A second arising question is therefore ``How do we design efficient relaying user scheduling protocols in the generalized $K$-user CRS?"


	\item Since only the two-user MISO BC is considered in \cite{Jian2019CRS}, the impact of the number of  users on the system performance remains unexplored. In the $K$-user case,  the search space of the optimal solution is enlarged since  the common rate is shared by more users and the relaying user scheduling becomes more complex as $K$ increases.  Hence, a third question arising is ``How will the CRS design, optimization and performance change with the number of  users $K$?"
\end{itemize}
 In this work,  we derive a generalized $K$-user CRS framework and provide solutions to overcome the above issues. The major contributions are summarized as follows: 

\begin{itemize}
\item We formulate a  max-min fairness problem  of a $K$-user CRS network in order to optimize the precoders, message split and time slot allocation as well as relaying user scheduling.  The objective is to maximize the minimum Quality of Service (QoS) rate among users subject to a  transmit power constraint. This is the first work in the literature on the design of the generalized $K$-user CRS.

\item We study both the centralized and decentralized  user scheduling for the $K$-user CRS.  Through analyzing the global optimal point of the max-min fairness problem, we prove that the  worst common rate  achieved by the relaying users in the optimal relaying user group is always larger than that achieved by other users in the direct transmission phase.  Motivated by the proposition, efficient centralized and decentralized relaying protocols are proposed and compared. Both protocols only require the information of channel strengths for relaying-user selection.

\item We propose a Successive Convex Approximation (SCA)-based algorithm to jointly optimize the precoders, message split and time slot allocation. The qualitative complexities of the proposed algorithm and the algorithm proposed in \cite{Jian2019CRS} are analyzed. The proposed SCA-based algorithm dramatically reduces the computational complexity since only a single layer iterative procedure is required to yield a solution compared with the two-layer iterative procedures in \cite{Jian2019CRS}.

\item We show through numerical results that the proposed relaying protocols with one single relaying user selection is able to achieve a rate performance very close to that of the optimal relaying scheduling algorithm in a wide range of network loads (underloaded and overloaded regimes), user deployments (with a diversity of channel strengths) and  power constraints at the relaying users. 

\item We further demonstrate that  the proposed SCA-based algorithm achieves the same  performance as the Weighted Minimum Mean Square Error (WMMSE) and exhaustive search-based algorithm proposed in \cite{Jian2019CRS} while the computational complexity of the system is much reduced.  Moreover, we observe that the performance benefits of CRS over Non-cooperative RS (NRS) become more significant as the number of transmit antennas decreases or as  the number of users/the channel strength disparity among user increases.  The  observations  bring the conclusion that by letting the user with the strongest channel strength forward the common stream to other users with weaker channel strengths, the rate of decoding the common stream at all users can be further enhanced  and therefore, the worst-case achievable rate of using CRS is further boosted even at low SNR.
\end{itemize}

\subsection{Organization}
The rest of the paper is organized as follows.
In Section \ref{sec: system model}, we introduce the system model of $K$-user CRS. The  max-min fairness optimization problem is formulated in Section \ref{sec: problem formulation}. The proposed low-complexity  relaying protocol as well as the SCA-based joint optimization algorithm are specified in Section \ref{sec: proposed algorithm}. Numerical results illustrating the benefits of the proposed algorithm are discussed in Section \ref{sec: numerical results}, followed by the conclusions in Section \ref{sec: conclusion}.

\subsection{Notations} 
The superscript $(\cdot)^T$  denotes transpose and $(\cdot )^H$ denotes conjugate-transpose.
 $\mathcal{CN}(\delta ,\sigma^2)$ represents a complex Gaussian distribution with mean $\delta $ and variance $\sigma^2$.
The boldface uppercases represent matrices and lowercase letters represent vectors.
$\mathrm{tr}(\cdot)$ is the trace. $\left\vert\cdot\right\vert$ is the absolute value and
$\left\Vert\cdot\right\Vert$ is the Euclidean norm. $\mathbb{C}$ denotes the complex space.   $|\mathcal{A}|$ is the cardinality of the set $\mathcal{A}$.

\section{System Model}
\label{sec: system model}

\begin{figure}
	\centering
	\includegraphics[width=3.3in]{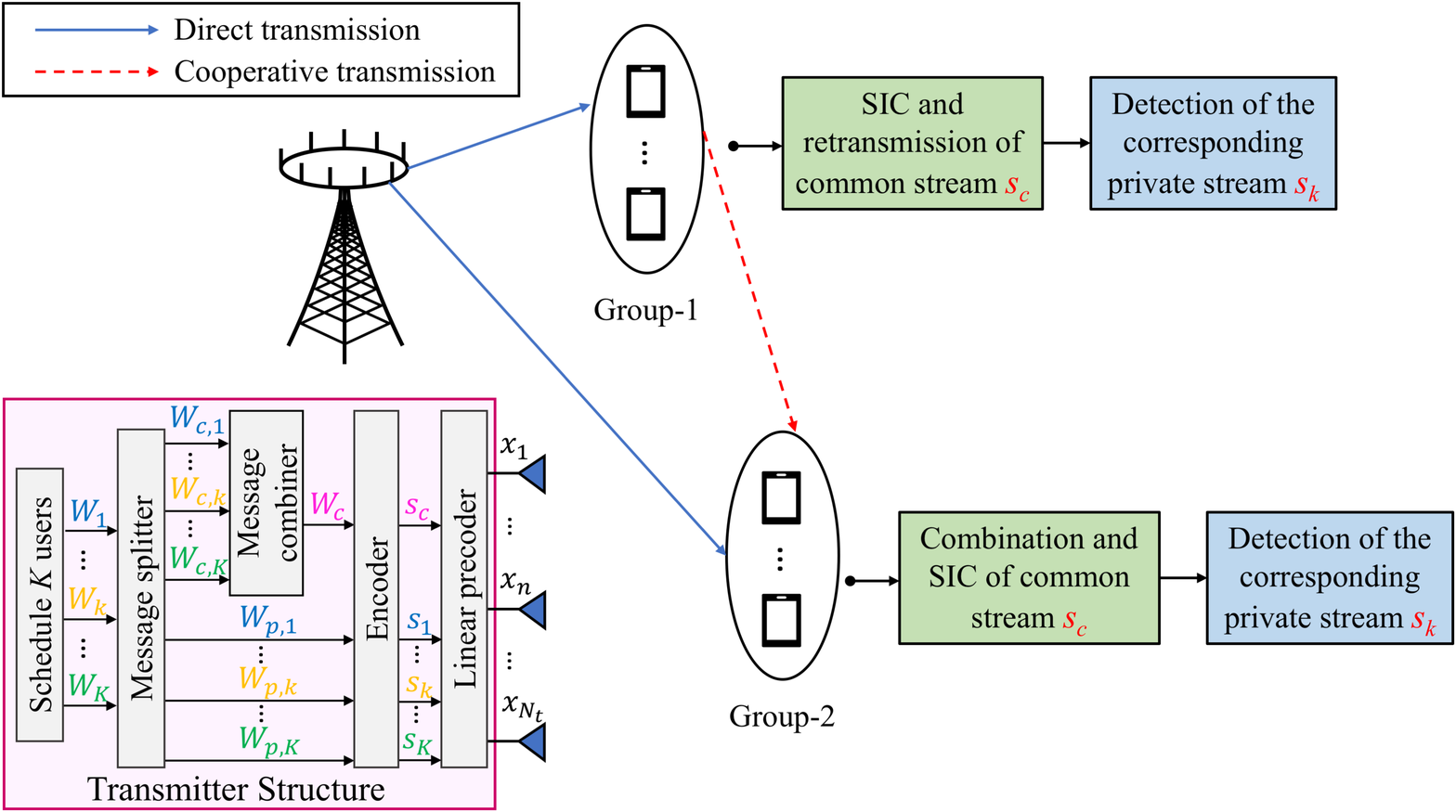}
	\caption{The proposed $K$-user cooperative rate-splitting system.}
	\label{fig: systemModel}
\end{figure}

The system model is illustrated in Fig. \ref{fig: systemModel}.  BS equipped with $N_t$ transmit antennas simultaneously serves $K$ single antenna users indexed by the set $\mathcal{K}=\{1,2,\ldots,K\}$. The users are divided into two separated user groups, namely, group-1 indexed by the set $\mathcal{K}_1$  and group-2 indexed by the set   $\mathcal{K}_2$, where $\mathcal{K}_1\cup \mathcal{K}_2=\mathcal{K}$.

\begin{figure}
	\centering
	\includegraphics[width=3.5in]{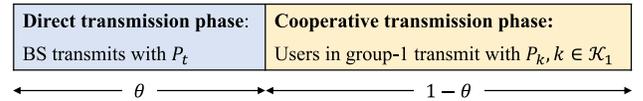}
	\caption{Time slot allocation for the two transmission phases. }
	\label{fig: timeSlot}
\end{figure}

The signal transmission is accomplished in the direct transmission phase and cooperative transmission phase. As illustrated in Fig. \ref{fig: timeSlot}, the time slot allocation for the two phases may not  be equal. In the first time slot (also known as direct transmission phase),  BS transmits signals to all  users. $\theta$ is the fraction of time allocated to the direct transmission phase. In the second time slot (also known as the cooperative transmission phase), the users in group-1 cooperatively forward the signals to the users in group-2. The detailed transmission model is explained in the following subsections of the direct transmission phase and the cooperative transmission phase, respectively. 

\subsection{Direct transmission phase}

 In the direct transmission phase,  BS serves $K$ users as the typical  multi-antenna RS  transmission investigated in the recent literature \cite{RS2015bruno,RS2016hamdi,RSintro16bruno,mao2017rate,bruno2019wcl}. Following the principle of RS, the message $W_k$ intended for user-$k$ is split into a common part $W_{c,k}$ and a private part $W_{p,k}$. The common parts of all users $W_{c,1},\ldots,W_{c,K}$ are combined into a common message $W_c$. It is encoded into the common stream $s_c$ using a codebook shared by all users. $s_c$ should be decoded by all users as it contains part of the messages of all users. The private part $W_{p,k}$  of user-$k$ is independently encoded into the private stream $s_k$, which is only required to be decoded by user-$k$. The encoded streams are linearly precoded and the resulting transmit signal at BS in the first time slot is
 \begin{equation}
 	\mathbf{x}^{[1]}=\mathbf{P}\mathbf{s}=\mathbf{p}_cs_c+\sum_{k\in\mathcal{K}}\mathbf{p}_ks_k.
 \end{equation}
 Denoting $\mathbf{s}=[s_c, s_1, s_2, \ldots, s_K]^T$ and assuming that $\mathbb{E}[\mathbf{s}\mathbf{s}^H]=\mathbf{I}$, the average transmit power constraint is given by $\textrm{tr}(\mathbf{P}\mathbf{P}^H)\leq P_t$, where $\mathbf{P}=[\mathbf{p}_c, \mathbf{p}_1, \mathbf{p}_2,\ldots,\mathbf{p}_K ]$ is the integrated precoder matrix and $\mathbf{p}_c, \mathbf{p}_k \in\mathbb{C}^{N_t\times 1}$. 
  
The received signal at user-$k$ in the first time slot is given by
  \begin{equation}
  y_{k}^{[1]}=\mathbf{h}_k^H\mathbf{x}^{[1]}+n_k, \forall k \in\mathcal{K},
  \end{equation}
 where  $\mathbf{h}_k\in\mathbb{C}^{N_t\times1}$ is the channel between BS and user-$k$. We assume that the Channel State Information (CSI) at the transmitter and receivers are perfect\footnote{The study of Non-cooperative RS (NRS) in imperfect CSIT has been widely studied in the literature \cite{ RS2016hamdi,RSintro16bruno,RS2016joudeh,chenxi2017brunotopology,Medra2018SPAWC,Lu2018MMSERS}. Due to the space limitations, the performance of CRS in imperfect CSIT is not investigated in this work. It is an interesting  problem for future work.}. $n_k\sim  \mathcal{CN}(0, 1) $ is the Additive White Gaussian Noise (AWGN). 
 Therefore, the transmit SNR is equal to the total power consumption $P_t$.

Users of group-1 first decode the common stream $s_c$ by fully treating the interference from all the private streams as noise. With the assist of SIC, the decoded common stream is removed from the received signal and then each user decodes the intended private stream by treating the remaining interference from other private streams as noise. Under the assumption of perfect SIC\footnote{The impact of imperfect SIC on the system performance is beyond the scope of this work, which we will investigate in our future work.}, the rate of decoding the common stream at each user of group-1 in the first time slot  is given by
   \begin{equation}
 R_{c,k}^{[1]}=\theta\log_2\left(1+\frac{\left|\mathbf{h}_k^H\mathbf{p}_c\right|^2}{\sum_{j\in\mathcal{K}}\left|\mathbf{h}_k^H\mathbf{p}_j\right|^2+1}\right),\forall k \in\mathcal{K}_1.
 \end{equation}
 The rate of decoding the private stream at each user of group-1  in the first time slot   is given by
 \begin{equation}
 R_{k}^{[1]}=\theta\log_2\left(1+\frac{\left|\mathbf{h}_k^H\mathbf{p}_k\right|^2}{\sum_{j\in\mathcal{K},j\neq k}\left|\mathbf{h}_k^H\mathbf{p}_j\right|^2+1}\right),\forall k \in\mathcal{K}_1.
 \end{equation}

 \subsection{Cooperative transmission phase}
 
 In the cooperative transmission phase,  users in group-1 employ the Non-regenerative Decode-and-Forward
(NDF) protocol \cite{TCOM2008NDF,mesbah2008power,NDF2010ICC} to forward the common stream $s_c$ to the users in group-2. Each user-$j$ of group-1 re-encodes $s_c$ with a  codebook generated independently from that of  BS and  forwards  $s_c$ to the users of group-2 with the transmit power $P_j$.  The transmit signal at user-$j$ of group-1 in the second time slot is given by
  \begin{equation}
x_{j}^{[2]}=\sqrt{  P_{j}}s_c, \forall j\in\mathcal{K}_1.
 \end{equation}

The signal received at user-$k$ of group-2  in the second time slot is 
  \begin{equation}
y_{k}^{[2]} = \sum_{j\in\mathcal{K}_1}h_{k,j}x_{j}^{[2]}+n_k, \forall k\in\mathcal{K}_2,
\end{equation}
where $h_{k,j}$ is the Single-Input Single-Output (SISO) channel between user-$k$ and user-$j$.

The rates of decoding the common stream at user-$k$ of group-2 in the first and the second time slots  are respectively given by
   \begin{equation}
R_{c,k}^{[1]}=\theta\log_2\left(1+\frac{\left|\mathbf{h}_k^H\mathbf{p}_c\right|^2}{\sum_{j\in\mathcal{K}}\left|\mathbf{h}_k^H\mathbf{p}_j\right|^2+1}\right),\forall k \in\mathcal{K}_2.
\end{equation}

\begin{equation}
R_{c,k}^{[2]}=(1-\theta)\log_2\left(1+\sum_{j\in\mathcal{K}_1}P_{j}\left|{h}_{k,j}^H\right|^2\right),\forall k \in\mathcal{K}_2.
\end{equation}
Users in group-2 combine the decoded common stream in both time slots. The achievable rate of decoding the common stream at all users with the NDF \cite{dynamicTDM2007DG,dynamicTDM2012OA,dynamicTDM2017JW} protocol\footnote{Note that there are two different types of  DF relaying protocols, namely,  NDF  and  Regenerative DF (RDF). The major difference between NDF and RDF is that the former considers independent  codebooks  while the latter uses the same codebook at BS and the relaying users. In this work, we focus on the application of RS in the NDF relaying protocol. Readers are referred to \cite{TCOM2008NDF,mesbah2008power,NDF2010ICC} for more detailed comparison of NDF and RDF protocols.} is  given as
\begin{equation}
\label{eq: common rate}
\begin{aligned}
R_{c}&=\min\left(\{R_{c,k}^{[1]}|k\in\mathcal{K}_1\},\{R_{c,k}^{[1]}+R_{c,k}^{[2]}|k\in\mathcal{K}_2\}\right)\\
        &=\min\left\{R_{c,1},R_{c,2}\right\},
\end{aligned}
\end{equation}
where $R_{c,1}=\min_{k\in\mathcal{K}_1}\{R_{c,k}^{[1]}\}$ and $R_{c,2}=\min_{k\in\mathcal{K}_2}\{R_{c,k}^{[1]}+R_{c,k}^{[2]}\}$ are the achievable common rate of users in group-1 and group-2, respectively. $R_{c}$ ensures that all users are able to decode the common stream successfully. As $R_c$ is shared by all users for the transmission of the common stream $s_c$, it is equal to $\sum_{k\in\mathcal{K}}C_{k}= R_c$, where $C_k$ is the portion of $R_c$ transmitting $W_{c,k}$. Once $s_c$ is decoded and removed from the received signal, user-$k$ decodes the intended private stream. 
The rate of decoding the private stream at each user of group-2   follows equation (4).
Denote the common rate vector as  $\mathbf{c}=[C_{1},\ldots,C_{K}]$.  $\mathbf{c}$ is required to be optimized jointly with the precoder $\mathbf{P}$ in order to maximize the worst-cast achievable rate. The total achievable rate of user-$k$ is given by $R_{k,tot}=R_{k}^{[1]} +C_k,\forall k\in\mathcal{K}$.

\section{Problem Formulation}
\label{sec: problem formulation}
In this work, we emphasize the user fairness issue. The precoder $\mathbf{P}$, the RS message split via common rate allocation $\mathbf{c}$, the time slot allocation $\theta$ and the relaying user  selection scheme $\mathcal{K}_1$ are optimized   with the objective of maximizing the minimum user rate.
The problem of  maximizing the worst achievable  rate among users has been studied in the literature of MISO BC  \cite{maxminSE2008,beamformOpt2008} as well as in the literature of RS  without user-relaying \cite{RS2016joudeh,hamdi2017bruno}. Motivated by those studies, we resort to investigate the max-min fairness in NDF CRS assisted transmission.  The problem at hand  differs from conventional max-min  rate problems in two ways:
1) the time slot allocation and the relaying user selection  in CRS are considered as optimization variables,
2) the rate constraint of the common stream is different from conventional RS in MISO BC without user-relaying due to the introduced NDF CRS transmission model. 
 The   max-min  rate  problem for the $K$-user CRS   is formulated as
\begin{subequations}
	\label{eq: rs}
\begin{align}\max_{\mathbf{{P}}, \mathbf{c}, \theta, \mathcal{K}_1}&\min_{k\in\mathcal{K}} \,\,R_{k,tot} \label{o1}\\
	\mbox{s.t.}\quad
	&\,\, \sum_{k\in\mathcal{K}}C_{k}\leq R_c \label{c3}\\
	&\,\, \text{tr}(\mathbf{P}\mathbf{P}^{H})\leq P_{t} \label{c4}\\
		&\,\, C_{k}\geq 0,\forall k\in\mathcal{K} \label{c5}\\
		&\,\, \mathcal{K}_1\subset \mathcal{K},  \mathcal{K}_2=\mathcal{K}\setminus \mathcal{K}_1. \label{c6}
	\end{align}
\end{subequations}
As    $ \mathcal{K}_2$ is determined once $ \mathcal{K}_1$ is selected, i.e., $\mathcal{K}_2=\mathcal{K}\setminus \mathcal{K}_1$, only $ \mathcal{K}_1$ is considered as a variable.  Constraint (\ref{c3}) guarantees that  the common stream is successfully decoded by all  users. Constraint (\ref{c4}) is the transmit power constraint at BS.

The formulated joint user selection and resource allocation problem is a mixed integer non-convex optimization problem.  The optimal relaying user selection problem itself   is of high computational complexity, particularly when the number of users is large. Therefore, in this paper, we propose a two-stage low-complexity algorithm to solve the  problem. In summary,   the algorithm has the following two stages:

\textit{Stage 1: Relaying user selection protocol:} The two user groups $\mathcal{K}_1$ and $\mathcal{K}_2$ are determined by selecting the set of relaying users  $\mathcal{K}_1$ that optimize the max-min fairness of the CRS transmission strategy.  Both centralized and decentralized relaying user selection protocols are studied in this work. The former enables relaying user selection at BS while the latter allows users to accomplish the selection in a decentralized manner.

\textit{Stage 2: Joint precoder design and resource allocation:} Based on the proposed relaying user selection protocols, we jointly optimize the precoders, message split and time slot allocation by solving the following  max-min problem:
\begin{equation}
	\label{eq: rs step2}
	\begin{aligned}
	\max_{\mathbf{{P}}, \mathbf{c}, \theta}\,\,\,\,&\min_{k\in\mathcal{K}} \,\, R_{k,tot} \\
	\mbox{s.t.}\quad
	&\,\, \textrm{(\ref{c3}), (\ref{c4}), (\ref{c5})}.
	\end{aligned}
\end{equation}
The proposed algorithm will be discussed thoroughly  in the next section.

\section{Proposed low-complexity algorithm}
\label{sec: proposed algorithm}
In this section, we elaborate on the proposed two-stage low-complexity algorithm to solve problem (\ref{eq: rs}).
\subsection{Stage 1: Relaying user selection protocol}

The transmission in the cooperative transmission phase helps to enhance the common rate $R_c$ since only the common stream $s_c$ is transmitted in this phase. As all users are required to decode $s_c$ in the direct transmission phase, they all have the ability of re-transmitting the common stream.
 Two questions arises when designing the  relaying user selection protocol: ``how do we select each relaying user?" and ``how many relaying users do we need?" To gain more insights, we further investigate the global optimal point $(\mathbf{{P}}^{\star}, \mathbf{c}^{\star}, \theta^{\star}, \mathcal{K}_1^{\star})$ of problem  (\ref{eq: rs}) and we obtain Proposition \ref{prop: relaySelection}:

\begin{prop} 
	\label{prop: relaySelection}
	 At the global optimal point $(\mathbf{{P}}^{\star}, \mathbf{c}^{\star}, \theta^{\star}, \mathcal{K}_1^{\star})$ of  problem (\ref{eq: rs}), the common rates achieved by the users in group-1 $R_{c,1}(\mathbf{{P}}^{\star},  \theta^{\star},\mathcal{K}_1^{\star})$ and group-2 $R_{c,2}(\mathbf{{P}}^{\star},  \theta^{\star},\mathcal{K}_2^{\star})$ are equal, which can be  mathematically written as
	\begin{equation}
	\label{eq: proposition 1.1}
	\begin{aligned}
	&\min_{k\in\mathcal{K}_1^{\star}}\{R_{c,k}^{[1]}(\mathbf{{P}}^{\star},\theta^{\star})\}\\=&\min_{k\in\{\mathcal{K}\setminus\mathcal{K}_1^{\star}\}}\{R_{c,k}^{[1]}(\mathbf{{P}}^{\star},  \theta^{\star})+R_{c,k}^{[2]}( \theta^{\star},\mathcal{K}_1^{\star})\},
	\end{aligned}
	\end{equation}
where 
\[
\resizebox{.98\width}{!} {$\begin{aligned}
R_{c,1}(\mathbf{{P}}^{\star},  \theta^{\star},\mathcal{K}_1^{\star})&=\min_{k\in\mathcal{K}_1^{\star}}\{R_{c,k}^{[1]}(\mathbf{{P}}^{\star},\theta^{\star})\},
\\R_{c,2}(\mathbf{{P}}^{\star},  \theta^{\star},\mathcal{K}_1^{\star})&=\min_{k\in\{\mathcal{K}\setminus\mathcal{K}_1^{\star}\}}\{R_{c,k}^{[1]}(\mathbf{{P}}^{\star}, \theta^{\star})+R_{c,k}^{[2]}( \theta^{\star},\mathcal{K}_1^{\star})\},
\end{aligned}$}\]
and the optimal relaying user grouping obeys the following rule when $0<\theta^{\star}<1$:
	\begin{equation}
	\label{eq: proposition 1.2}
	\min_{k\in\mathcal{K}_1^{\star}}\{R_{c,k}^{[1]}(\mathbf{{P}}^{\star},  \theta^{\star})\}>\min_{k\in\mathcal{K}\setminus\mathcal{K}_1^{\star}}\{R_{c,k}^{[1]}(\mathbf{{P}}^{\star},  \theta^{\star})\}.
	\end{equation} 
\end{prop}

\textit{Proof}:  We first prove the equivalence of the common rate achieved by users in group-1 and group-2  at the global optimal point  (\ref{eq: proposition 1.1}) by contradiction. By assuming $\hat{k}_1$ and $\hat{k}_2$ are the two users that respectively achieve the worst common rate in group-1 and group-2 at the global optimal  point $(\mathbf{{P}}^{\star}, \mathbf{c}^{\star}, \theta^{\star}, \mathcal{K}_1^{\star})$, we obtain that
\begin{equation}
\begin{aligned}
R_{c,1}(\mathbf{{P}}^{\star},  &\theta^{\star},\mathcal{K}_1^{\star})=\theta^{\star}f_{c,\hat{k}_1}^{[1]}(\mathbf{{P}}^{\star}),\\
R_{c,2}(\mathbf{{P}}^{\star},  &\theta^{\star},\mathcal{K}_1^{\star})=\theta^{\star}f_{c,\hat{k}_2}^{[1]}(\mathbf{{P}}^{\star})+(1-\theta^{\star})f_{c,\hat{k}_2}^{[2]}(\mathcal{K}_1^{\star})\\
&=\theta^{\star}\left(f_{c,\hat{k}_2}^{[1]}(\mathbf{{P}}^{\star})-f_{c,\hat{k}_2}^{[2]}(\mathcal{K}_1^{\star})\right)+f_{c,\hat{k}_2}^{[2]}(\mathcal{K}_1^{\star}), 
\end{aligned}
\end{equation}
where
 \[
\begin{aligned}
f_{c,k}^{[1]}(\mathbf{{P}}^{\star})&=\log_2\left(1+\frac{\left|\mathbf{h}_k^H\mathbf{p}_c^{\star}\right|^2}{\sum_{j\in\mathcal{K}}\left|\mathbf{h}_k^H\mathbf{p}_j^{\star}\right|^2+1}\right)\\
f_{c,k}^{[2]}(\mathcal{K}_1^{\star})&=\log_2\left(1+\sum_{j\in\mathcal{K}_1^{\star}}P_{j}\left|{h}_{k,j}^H\right|^2\right).
\end{aligned}
\]
Note that $f_{c,\hat{k}_2}^{[1]}(\mathbf{{P}}^{\star})<f_{c,\hat{k}_2}^{[2]}(\mathcal{K}_1^{\star})$ should hold when $0<\theta^{\star}<1$. Otherwise, both $R_{c,1}(\mathbf{{P}}^{\star},  \theta,\mathcal{K}_1^{\star})$ and $R_{c,2}(\mathbf{{P}}^{\star},  \theta,\mathcal{K}_1^{\star})$ are increasing functions of $\theta$ for the optimal $\mathbf{{P}}^{\star},\mathcal{K}_1^{\star}$ and the optimal $\theta^{\star}$ should be 1 in order to maximize the minimum rate among users. Hence, when $0<\theta<1$, $R_{c,1}(\mathbf{{P}}^{\star},  \theta,\mathcal{K}_1^{\star})$ is a monotonic increasing function of $\theta$  while $R_{c,2}(\mathbf{{P}}^{\star},  \theta,\mathcal{K}_1^{\star})$ is a monotonic decreasing function of $\theta$.

 If $R_{c,1}(\mathbf{{P}}^{\star},  \theta^{\star},\mathcal{K}_1^{\star})>R_{c,2}(\mathbf{{P}}^{\star},  \theta^{\star},\mathcal{K}_1^{\star})$, we have 
\begin{equation}
\resizebox{.92\width}{!} {$\begin{aligned}
&\theta^{\star}f_{c,\hat{k}_1}^{[1]}(\mathbf{{P}}^{\star})>\theta^{\star}f_{c,\hat{k}_2}^{[1]}(\mathbf{{P}}^{\star})+(1-\theta^{\star})f_{c,\hat{k}_2}^{[2]}(\mathcal{K}_1^{\star})\\\Leftrightarrow\,\,
 & \theta^{\star}>\frac{f_{c,\hat{k}_2}^{[2]}(\mathcal{K}_1^{\star})}{f_{c,\hat{k}_1}^{[1]}(\mathbf{{P}}^{\star})-f_{c,\hat{k}_2}^{[1]}(\mathbf{{P}}^{\star})+f_{c,\hat{k}_2}^{[2]}(\mathcal{K}_1^{\star})}\triangleq \Gamma(\mathbf{{P}}^{\star},\mathcal{K}_1^{\star}).
	\end{aligned}$}
\end{equation}
By decreasing $\theta$ from $\theta^{\star}$ to $\theta'=\Gamma(\mathbf{{P}}^{\star},\mathcal{K}_1^{\star})$,  we have $R_{c,2}(\mathbf{{P}}^{\star},  \theta',\mathcal{K}_1^{\star})>R_{c,2}(\mathbf{{P}}^{\star},  \theta^{\star},\mathcal{K}_1^{\star})$ and $R_{c,1}(\mathbf{{P}}^{\star},  \theta',\mathcal{K}_1^{\star})=R_{c,2}(\mathbf{{P}}^{\star},  \theta',\mathcal{K}_1^{\star})$. The achievable  common rate  $\sum_{k\in\mathcal{K}}C_{k}'$$=\min\{R_{c,1}(\mathbf{{P}}^{\star},  \theta',\mathcal{K}_1^{\star})$ $,R_{c,2}(\mathbf{{P}}^{\star},  \theta',\mathcal{K}_1^{\star})\}$ increases. By allocating the improved common rate to the worst-case user, the  value of the objective function achieved by using  the new solution $(\mathbf{{P}}^{\star}, \mathbf{c}', \theta',\mathcal{K}_1^{\star})$ is higher than  that of $(\mathbf{{P}}^{\star}, \mathbf{c}^{\star}, \theta^{\star},\mathcal{K}_1^{\star})$, which contradicts the fact that
$(\mathbf{{P}}^{\star}, \mathbf{c}^{\star}, \theta^{\star},\mathcal{K}_1^{\star})$ is the global optimal  point. Similarly, we obtain that if $R_{c,1}(\mathbf{{P}}^{\star},  \theta^{\star},\mathcal{K}_1^{\star})<R_{c,2}(\mathbf{{P}}^{\star},  \theta^{\star},\mathcal{K}_1^{\star})$, $\theta^{\star}< \Gamma(\mathbf{{P}}^{\star},\mathcal{K}_1^{\star})$ holds. A better solution is obtained by increasing $\theta$ from $\theta^{\star}$ to $\theta'=\Gamma(\mathbf{{P}}^{\star},\mathcal{K}_1^{\star})$. The contradiction arises. Hence, we draw the conclusion that at  the global optimal point $(\mathbf{{P}}^{\star}, \mathbf{c}^{\star}, \theta^{\star}, \mathcal{K}_1^{\star})$ of  problem (\ref{eq: rs}), $R_{c,1}(\mathbf{{P}}^{\star},  \theta^{\star},\mathcal{K}_1^{\star})=R_{c,2}(\mathbf{{P}}^{\star},  \theta^{\star},\mathcal{K}_1^{\star})$. 

As $\min_{k\in\{\mathcal{K}\setminus\mathcal{K}_1^{\star}\}}\{R_{c,k}^{[1]}(\mathbf{{P}}^{\star},  \theta^{\star})+R_{c,k}^{[2]}( \theta^{\star},\mathcal{K}_1^{\star})\}$
$\geq\min_{k\in\{\mathcal{K}\setminus\mathcal{K}_1^{\star}\}}\{R_{c,k}^{[1]}(\mathbf{{P}}^{\star},  \theta^{\star})\}+\min_{k\in\{\mathcal{K}\setminus\mathcal{K}_1^{\star}\}}\{R_{c,k}^{[2]}( \theta^{\star},\mathcal{K}_1^{\star})\}$ and $R_{c,k}^{[2]}( \theta^{\star},\mathcal{K}_1^{\star})>0, \forall k\in\mathcal{K}\setminus\mathcal{K}_1^{\star}$, we obtain that
$R_{c,2}(\mathbf{{P}}^{\star},  \theta^{\star},\mathcal{K}_1^{\star}$ $)>\min_{k\in\{\mathcal{K}\setminus\mathcal{K}_1^{\star}\}}\{R_{c,k}^{[1]}(\mathbf{{P}}^{\star},$ $  \theta^{\star})\}$. Based on  (\ref{eq: proposition 1.1}), we have $R_{c,1}(\mathbf{{P}}^{\star},  \theta^{\star},\mathcal{K}_1^{\star})>\min_{k\in\{\mathcal{K}\setminus\mathcal{K}_1^{\star}\}}\{$ $ R_{c,k}^{[1]}(\mathbf{{P}}^{\star}, \theta^{\star})\}$ when $0<\theta^{\star}<1$. Proof of proposition \ref{prop: relaySelection} is completed.
\qed

\begin{figure}
	\centering
	\includegraphics[width=3.5in]{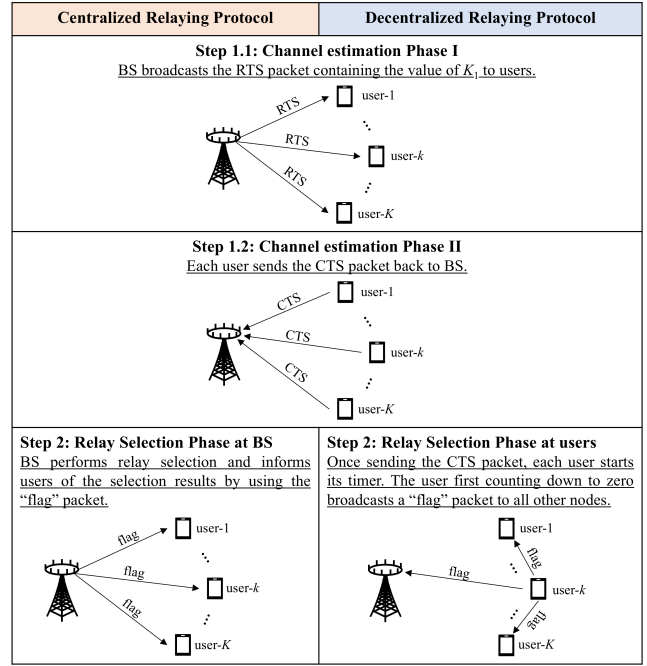}
	\caption{The proposed centralized and decentralized relaying  user selection protocols, $K_1=1$.}
	\label{fig: relayingProtocal}
\end{figure}

Proposition 1 provides insights into the design of relaying user selection protocol. To boost the common rate via CRS,  users with large $R_{c,k}^{[1]}$ are suggested to be clustered in $\mathcal{K}_1$ while the users with low $R_{c,k}^{[1]}$ are suggested to be clustered in $\mathcal{K}_2$\footnote{Though Proposition \ref{prop: relaySelection} only provides the selection policy for the worst-case users in $\mathcal{K}_1$ and $\mathcal{K}_2$, we generalize it to all users in the two groups. There are two major reasons based on the analysis of Karush–Kuhn–Tucker  (KKT) conditions  as well as the computational complexity of problem (\ref{eq: rs}).  Mathematically, to reach the KKT point of problem (\ref{eq: rs}), the complementary conditions should be met. Combining constraint (\ref{c3}) and Proposition \ref{prop: relaySelection}, we obtain the following constraint: $\mu_k(\sum_{k\in\mathcal{K}}C_{k}-R_{c,k}^{[1]})=0,\forall k\in\mathcal{K}_1$, where $\mu_k\geq 0$ is the Lagrangian dual variable. When all common  rate constraints are active ($\mu_k> 0$), we obtain that $\sum_{k\in\mathcal{K}}C_{k}-R_{c,k}^{[1]}=0$ holds for all users in $\mathcal{K}_1$. Similarly, we could derive the same conclusion for users in group-2. Hence, all users in group-1 achieve the same optimal common rate, which is larger than the common rate achieved by all users in group-2. When some of the common rate constraints are inactive, the problem becomes much difficult to handle. The computational complexity of  optimizing relaying-user scheduling is NP-hard already. In total, $\sum_{k=1}^{K-1}\binom{K}{k}$ possible relaying-user groups should be searched. For each grouping method, the precoders, message split and time slot allocation are required to be optimized so as to obtain the achievable max-min rate for comparison. To gain insights in such circumstance is hard. }. As $s_c$ is decoded by all users in the first time slot,  the rate of $R_{c,k}^{[1]}$ is strongly influenced by  users' channel strengths $\left\Vert\mathbf{h}_k\right\Vert^2, k\in\mathcal{K}$. Hence, an intuitive and simple selection algorithm is based on the channel strengths of users. In this work, we extend the Request-To-Send (RTS)/Clear-To-Send (CTS)  collision avoidance mechanism proposed in \cite{relaySelection2006RTS} to our $K$-user dynamic CRS network.  In principle, by exchanging  RTS and CTS packets between BS and  users, each user can not only estimate the channel strength in between but also get synchronized after handshake \cite{relaySelection2006RTS}.  Both centralized and decentralized relaying protocols based on users' channel strengths  are studied. Fig. \ref{fig: relayingProtocal} illustrates the  two proposed protocols when a single "best" relaying user is  selected ($K_1=1$). 

 \subsubsection{Centralized relaying protocol} As BS requires the CSI to design the precoders, one natural method is to select  relaying users centrally at BS.  The proposed centralized relaying protocol is summarized in Algorithm \ref{algo: centralized}. Once BS obtains the CSI of all users after handshake in Step 1, it selects the $K_1$ users with highest channel gain in Step 2.  Once the $K_1$ users are selected, BS encapsulates the selection result into one ``flag" packet and broadcasts it to all users. Each user decodes the received packet and gets the its own selection result.
\begin{algorithm}[t!]
	\textbf{Initialize:} $n\leftarrow0, \mathcal{K}_1=\emptyset, \mathcal{K}_2=\mathcal{K}$\;
	
	\textbf{Step 1:}  BS broadcasts RTS packet to all users. Each user receives $K_1$, replies BS with CTS.  BS estimates the channel $\mathbf{h}_k$ and calculates the channel strength $\left\Vert\mathbf{h}_k\right\Vert^2$\;

	\textbf{Step 2:}  BS selects $K_1$ users with highest channel gain as \\
	\Repeat{$|\mathcal{K}_1|=K_1$}{
		 $k^{\star}=\max_{k}\{\left\Vert\mathbf{h}_k\right\Vert^2\}$\;
		 $\mathcal{K}_1\leftarrow\mathcal{K}_1\cup\{k^{\star}\}$\; $\mathcal{K}_2\leftarrow\mathcal{K}_2\setminus \{k^{\star}\}$\;		
	}
\textbf{Step 3:}  BS encapsulates the selection results into  one ``flag"  packet and broadcasts it to all users.\\
	\caption{Proposed centralized relaying protocol}
	\label{algo: centralized}		
\end{algorithm}

\subsubsection{Decentralized relaying protocol} 
 As all users are capable of serving as relays, the cooperative transmission networks can be designed adaptively based on the CSI conditions. However,  due to the fact that the RTS packet is sent through a common pilot channel shared by all users while the CTS packets are fed back through uplink pilot channels dedicated to the corresponding users,  it is easier to achieve perfect CSIR  obtained from the common pilot compared with CSIT obtained from the uplink dedicated pilot \cite{TWC2008Lau}. Hence, decentralized selection approaches are preferred as they are not influenced by the CSIT inaccuracy.

The proposed  decentralized relaying protocol is summarized in Algorithm \ref{algo: decentralized}.  After handshake in step 1, each UE knows the number of relaying users $K_1$ to select as well as  the channel strength $\left\Vert\mathbf{h}_k\right\Vert^2$. Perfect synchronization among users and BS is assumed after step 1.  In the second step, each user starts the relaying user selection procedure. With the assistance of timer $T_k=\frac{\lambda}{\left\Vert\mathbf{h}_k\right\Vert^2}, \forall k\in\mathcal{K}_2$, where $\lambda$ is a constant in time units \cite{relaySelection2006RTS},  all users start counting down.  The user first counting down to zero broadcasts a ``flag" packet to all other users to announce itself as the best relaying user. All other users add it to their relaying user list and reset the timer  for the next round selection. The selection procedure repeats until $K_1$ relaying users are selected. 
As ``how many relaying users do we need?" is a critical issue, the number of relaying users $K_1$ is considered to be a variable in the proposed algorithm. The influence of $K_1$ will be investigated in the numerical results section. The simplest selection is when $K_1=1$, which is a single ``best" relaying user selection. In the numerical results, we will show that $K_1=1$ achieves a rate almost the same as the rate achieved by the optimal relaying user scheduling.

\textit{Remark 1: Compared with the decentralized relaying protocol without RS in \cite{relaySelection2006RTS}, the protocol proposed for RS-assisted transmission is simplified as  only the CSI between BS and the user is required at each user to perform decentralized relaying user selection. In contrast, without RS, each user should obtain CSI between every other user and itself as well as the CSI between BS and itself to perform decentralized relaying user selection \cite{relaySelection2006RTS}. Moreover, the protocol proposed in  \cite{relaySelection2006RTS} can only select a single relaying user while our proposed protocol is generalized for multi-relaying user selection. }

\begin{algorithm}[t!]
	\textbf{Initialize:} $n\leftarrow0, \mathcal{K}_1=\emptyset, \mathcal{K}_2=\mathcal{K}$\;
	
	\textbf{Step 1:} BS broadcasts RTS packet that contains the value of $K_1$ to all users. Each user receives $K_1$, replies BS with CTS, estimates the channel $\mathbf{h}_k$ and calculates the channel strength $\left\Vert\mathbf{h}_k\right\Vert^2$\;

	\textbf{Step 2:} Decentralized relaying user selection:\\
	\Repeat{$|\mathcal{K}_1|=K_1$}{
		At each user in $\mathcal{K}_2$, clear the existing timer if it has one\;
		The timer is reset at each user in $\mathcal{K}_2$ as $T_k=\frac{\lambda}{\left\Vert\mathbf{h}_k\right\Vert^2}, \forall k\in\mathcal{K}_2$\;
		The user first counts down to zero, broadcasts a flag packet to all users in $\mathcal{K}_2$ to announce itself as the best relaying user in $\mathcal{K}_2$\; 
		Once users in $\mathcal{K}_2$ receive the ``flag" packet from the best relaying user $k^{\star}$, each of them update $\mathcal{K}_1\leftarrow\mathcal{K}_1\cup\{k^{\star}\}$, $\mathcal{K}_2\leftarrow\mathcal{K}_2\setminus \{k^{\star}\}$\;		
	}
	\caption{Proposed decentralized relaying protocol}
	\label{algo: decentralized}		
\end{algorithm}

\subsubsection{Comparison} 
As illustrated in Fig. \ref{fig: relayingProtocal}, the major difference of the two proposed  protocols stems from the relay selection phase where the nodes performing relay selection and broadcasting the ``flag" packet are different. We further compare the proposed protocols in terms of overhead, synchronization and CSI requirement. 

Table \ref{tab: overhead} illustrates the overhead comparison in terms of signaling and time consumption between the two proposed relaying protocols as well as the strategies without user relaying. $N_{\textit{RTS}}$ and $N_{\textit{CTS}}$ represent  the number of symbols in each RTS and CTS packet, respectively. $T_{\textit{S}}$ denotes the symbol duration. $N_{\textit{flag}}^{C}$ and $N_{\textit{flag}}^{D}$  are the number of symbols in each flag packet of the centralized and  decentralized relaying protocols, respectively. Note that $N_{\textit{flag}}^{C}$  is  larger than $N_{\textit{flag}}^{D}$ (i.e.,$ N_{\textit{flag}}^{C}>N_{\textit{flag}}^{D}$) since the ``flag" packet sent by BS in the centralized relaying protocol contains the selection information of all users while the one sent by the selected user  in the  decentralized relaying protocol  only contains the selection result of one single user. 
 The handshake based on RTS and CTS is required for all strategies so as to obtain CSI at BS. Therefore,  the signaling overhead of RTS $N_{\textit{RTS}}$ and CTS for $K$ users $N_{\textit{CTS}}K$ (assuming TDD system) is required for all strategies. 
Comparing the signaling and time overhead of all strategies, the signaling  overhead of the centralized relaying protocol is relatively higher since $N_{\textit{flag}}^{C}=KN_{\textit{flag}}^{D}$  when BS simultaneously sends $K$ ``flag" packets to all users (via spatial multiplexing). In such case, the size of each ``flag" packet  for the centralized protocol is the same as the one for the decentralized protocol. 
However, the time consumption of the decentralized relaying protocol is higher since additional timer is required for the selection of each relaying user. Apparently, the strategies without user relaying has the lowest overhead.

Synchronization is required to be more accurate in the two strategies with user relaying  compared with those without user relaying. In the relay selection phase, no explicit synchronization protocol is required among the users since the timing process is triggered by the CTS or flag packet and the collision due to the inaccurate synchronization can be controlled by $\lambda$ \cite{relaySelection2006RTS}. A larger $\lambda$ will reduce the probability of collision\footnote{ If the  transmission still fails for a large $\lambda$, one solution is to switch from the  decentralized relaying protocol to the centralized one.  The study on  the failure of transmission is beyond the research scope in this paper, which will be considered as a future research direction.}. Readers are referred to  \cite{relaySelection2006RTS} for more details. In the information transmission phases (as illustrated in Fig. \ref{fig: systemModel} and Fig. \ref{fig: timeSlot}), the synchronization is assumed to be perfect. This assumption is commonly made in the literature of multi-relay cooperative transmission \cite{CL2010synchronize,TWC2011synchronize,TCOM2016overhead}. In the numerical results, we have shown that a single relaying user selection is enough to achieve the near optimal result. Hence, synchronization among relaying users is not necessary.

As mentioned previously, the centralized relaying protocol requires BS to select relaying users based on the obtained CSIT while the decentralized relaying protocol allows users to select relaying users themselves based on CSIR. Hence, the major motivation of adopting the decentralized protocol is that its performance is not sensitive to the CSIT inaccuracy at all. In contrast, the performance of centralized relaying selection drops as CSIT becomes worse.

 \begin{table*}[t!]
	\centering
	\caption{{Overhead comparison of different strategies}}
	\label{tab: overhead}
	
		\begin{tabular}{|L{2.9cm}|L{3.4cm}|L{6cm}|}
	\hline
	\textbf{Schemes}        & \textbf{Signaling overhead} & \textbf{Time overhead} \\ \hline
	Centralized relaying   &      $N_{\textit{RTS}}+N_{\textit{CTS}}K+N_{\textit{flag}}^{C}$                        &    $N_{\textit{RTS}}T_{\textit{S}}+N_{\textit{CTS}}T_{\textit{S}}K+N_{\textit{flag}}^{C}T_{\textit{S}}$                    \\ \hline
	Distributed relaying   &     $N_{\textit{RTS}}+N_{\textit{CTS}}K+N_{\textit{flag}}^{D}K_1$                         &        $N_{\textit{RTS}}T_{\textit{S}}+N_{\textit{CTS}}T_{\textit{S}}K+N_{\textit{flag}}^{D}T_{\textit{S}}K_1+\sum_{k\in\mathcal{K}_1}T_k$                 \\ \hline
	Without user relaying&            $N_{\textit{RTS}}+N_{\textit{CTS}}K$                     &           $N_{\textit{RTS}}T_{\textit{S}}+N_{\textit{CTS}}T_{\textit{S}}K$             \\ \hline
\end{tabular}
\end{table*}

\subsection{Stage 2: Joint precoder design and resource allocation}
The two user groups $\mathcal{K}_1$ and $\mathcal{K}_2$ are determined after stage 1. 
In the second stage, a SCA-based algorithm is proposed to jointly optimize the precoders, message split and time resource allocation for  problem (\ref{eq: rs step2}). 
 We first equivalently transform (\ref{eq: rs step2}) by introducing an auxiliary variable $t$:
\begin{subequations}
	\label{eq: rs step2 eqv}
	\begin{align}\max_{\mathbf{{P}}, \mathbf{c}, \theta,t}\,\,\,\,&\,\,\,t \label{eqv step2 const:o1}\\
	\mbox{s.t.}\quad
	&\,\, R_{k}^{[1]} +C_k\geq t, \forall k\in\mathcal{K} \label{eqv step2 const:c1}\\
	&\,\, \textrm{(\ref{c3}), (\ref{c4}), (\ref{c5})}. \nonumber
	\end{align}
\end{subequations}
The equivalence between (\ref{eq: rs step2 eqv}) and (\ref{eq: rs step2}) is guaranteed by noting
that constraint (\ref{eqv step2 const:c1}) is the same as $\min_{k\in\mathcal{K}}(R_{k}^{[1]} +C_k)\geq t $ and it must hold with equality at optimum. The problem is still non-convex due to the rate expressions  $R_{k}^{[1]}$ and $R_{c,k}^{[1]}$ in (\ref{eqv step2 const:c1}) and (\ref{c3}). Next, we introduce slack variable vectors $\bm{\alpha}=[\alpha_1,\ldots,\alpha_K]$, $\bm{\alpha}_c=[\alpha_{c,1},\ldots,\alpha_{c,K}]$, $\bm{\rho}=[\rho_1,\ldots,\rho_K]$, $\bm{\rho}_c=[\rho_{c,1},\ldots,\rho_{c,K}]$.  $\bm{\alpha}$ and $\bm{\alpha}_c$ are adopted to respectively represent the rate vectors of the private streams  and the common streams  at all users. $\bm{\rho}$ and $\bm{\rho}_c$  respectively denote the Signal-to-Interference-plus-Noise Ratio (SINR) vectors of the private streams  and the common streams. 
With the assistance of the new variables, the optimization problem (\ref{eq: rs step2 eqv}) is equivalently rewritten into 
\begin{subequations}
	\label{eq: rs step2 eqv2}
	\begin{align}\max_{\substack{\mathbf{{P}}, \mathbf{c}, \theta,t, \\ \bm{\alpha},\bm{\alpha}_c,\bm{\rho},\bm{\rho}_c}}\,\,\,\,&\, \,\,t \label{eqv2 step2 const:o1}\\
	\mbox{s.t.}\quad
	&\,\, \theta\alpha_k+C_k\geq t,\forall k\in\mathcal{K}  \label{eqv2 step2 const:c1}\\
	&\,\, \theta\alpha_{c,k}\geq \sum_{j\in\mathcal{K}}C_{j},\forall k\in\mathcal{K}_1 \label{eqv2 step2 const:c2}\\
	&\,\,  \theta\alpha_{c,k}+R_{c,k}^{[2]}\geq \sum_{j\in\mathcal{K}}C_{j},\forall k\in\mathcal{K}_2 \label{eqv2 step2 const:c3}\\
	&\,\, 1+\rho_k-2^{\alpha_k}\geq 0,\forall k\in\mathcal{K}  \label{eqv2 step2 const:c5}\\
	&\,\, 1+\rho_{c,k}-2^{\alpha_{c,k}}\geq 0,\forall k\in\mathcal{K} \label{eqv2 step2 const:c6}\\
	&\,\, \frac{\left|\mathbf{h}_k^H\mathbf{p}_k\right|^2}{\sum_{j\in\mathcal{K},j\neq k}\left|\mathbf{h}_k^H\mathbf{p}_j\right|^2+1}\geq \rho_k,\forall k\in\mathcal{K} \label{eqv2 step2 const:c7}\\
	&\,\, \frac{\left|\mathbf{h}_k^H\mathbf{p}_c\right|^2}{\sum_{j\in\mathcal{K}}\left|\mathbf{h}_k^H\mathbf{p}_j\right|^2+1}\geq \rho_{c,k},\forall k\in\mathcal{K}\label{eqv2 step2 const:c8}\\
	&\,\, \textrm{(\ref{c4}), (\ref{c5})}. \nonumber                                                                                                                                                                                                                                                                                                                                                                                                                                                                               
	\end{align}
\end{subequations}
To deal with all the non-convex constraints (\ref{eqv2 step2 const:c1})--(\ref{eqv2 step2 const:c3}), (\ref{eqv2 step2 const:c7}), (\ref{eqv2 step2 const:c8}), we adopt the following SCA method.
For constraints (\ref{eqv2 step2 const:c1})--(\ref{eqv2 step2 const:c3}), the bilinear function $\theta\alpha_k$ can be equivalently written as $\theta\alpha_k=\frac{1}{4}(\theta+\alpha_k)^2-\frac{1}{4}(\theta-\alpha_k)^2$. Hence, $\theta\alpha_k$ is approximated at the point $(\theta^{[n]},\alpha_k^{[n]})$ by the first-order Taylor approximation of $(\theta+\alpha_k)^2$, which is given by
\begin{equation}
\begin{aligned}
\theta\alpha_k\geq \frac{1}{2}\left(\theta^{[n]}+\alpha_k^{[n]}\right)&\left(\theta+\alpha_k\right)-\frac{1}{4}\left(\theta^{[n]}+\alpha_k^{[n]}\right)^2\\
&-\frac{1}{4}\left(\theta-\alpha_k\right)^2\triangleq \Phi^{[n]}(\theta,\alpha_{k})
	\end{aligned}
\end{equation}
Constraints (\ref{eqv2 step2 const:c1})--(\ref{eqv2 step2 const:c3}) are approximated around the point $(\theta^{[n]},\bm{\alpha}^{[n]},\bm{\alpha}_c^{[n]})$ at iteration $n$  as
\begin{equation}
\label{eq: c1-c3}
\begin{aligned}
&\,\, \Phi^{[n]}(\theta,\alpha_{k})+C_k\geq t,\forall k\in\mathcal{K}, \\
&\,\, \Phi^{[n]}(\theta,\alpha_{c,k})\geq \sum_{j\in\mathcal{K}}C_{j},\forall k\in\mathcal{K}_1, \\
&\,\,  \Phi^{[n]}(\theta,\alpha_{c,k})+R_{c,k}^{[2]}\geq \sum_{j\in\mathcal{K}}C_{j},\forall k\in\mathcal{K}_2. 
\end{aligned}
\end{equation}
Constraints (\ref{eqv2 step2 const:c7}), (\ref{eqv2 step2 const:c8}) are equivalently written into Difference-of-Convex (DC) forms, which are given by
\begin{equation}
\label{eq: DC}
\begin{aligned}
\sum_{j\in\mathcal{K},j\neq k}|\mathbf{{h}}_{k}^{H}\mathbf{{p}}_{j}|^{2}+1-\frac{|\mathbf{{h}}_{k}^{H}\mathbf{{p}}_{k}|^{2}}{ \rho_k}&\leq 0,\forall k\in\mathcal{K},\\
\sum_{j\in\mathcal{K}}|\mathbf{{h}}_{k}^{H}\mathbf{{p}}_{j}|^{2}+1-\frac{|\mathbf{{h}}_{k}^{H}\mathbf{{p}}_{c}|^{2}}{\rho_{c,k}}&\leq 0,\forall k\in\mathcal{K}.
\end{aligned}
\end{equation}
By reconstructing the concave parts of the DC constraints with the first-order Taylor approximations, the constraints (\ref{eqv2 step2 const:c7}), (\ref{eqv2 step2 const:c8}) are respectively approximated  around the point $(\mathbf{{P}}^{[n]},\bm{\rho}^{[n]},\bm{\rho}_c^{[n]})$ at iteration $n$ by
\begin{equation}
\label{eq: DC approxi}
\begin{aligned}
\sum_{j\in\mathcal{K},j\neq k}|\mathbf{{h}}_{k}^{H}\mathbf{{p}}_{j}|^{2}+1-&\frac{2\Re\{(\mathbf{{p}}_{k}^{[n]})^H{\mathbf{h}}_{k}{\mathbf{h}}_{k}^{H}\mathbf{{p}}_{k}\}}{ \rho_k^{[n]}}\\&+\frac{|\mathbf{{h}}_{k}^{H}\mathbf{{p}}_{k}^{[n]}|^{2}\rho_k}{(\rho_k^{[n]})^2}\leq 0,\forall k\in\mathcal{K},\\
\sum_{j\in\mathcal{K}}|\mathbf{{h}}_{k}^{H}\mathbf{{p}}_{j}|^{2}+1-&\frac{2\Re\{(\mathbf{{p}}_{c}^{[n]})^H{\mathbf{h}}_{k}{\mathbf{h}}_{k}^{H}\mathbf{{p}}_{c}\}}{ \rho_{c,k}^{[n]}}\\&+\frac{|\mathbf{{h}}_{k}^{H}\mathbf{{p}}_{c}^{[n]}|^{2}\rho_{c,k}}{(\rho_{c,k}^{[n]})^2}\leq 0,\forall k\in\mathcal{K}.
\end{aligned}
\end{equation}
Based on the above approximation methods, the original optimization problem can be solved using the SCA method. The main idea of SCA is to successively solve a sequence of convex subproblems. At iteration $n$, based on the optimal solution $(\mathbf{{P}}^{[n]},\theta^{[n]}, \bm{\alpha}^{[n]},\bm{\alpha}_c^{[n]},\bm{\rho}^{[n]},\bm{\rho}_c^{[n]})$ obtained from the previous iteration $n-1$, we solve the following subproblem:
\begin{equation}
	\label{eq: rs step2 appro}
	\begin{aligned}
	\max_{\substack{\mathbf{{P}}, \mathbf{c}, \theta,t, \\ \bm{\alpha},\bm{\alpha}_c,\bm{\rho},\bm{\rho}_c}}\,\,\,\,&\, \,\,t \\
	\mbox{s.t.}\quad
	&\,\, \textrm{(\ref{c4}), (\ref{c5}), (\ref{eqv2 step2 const:c5}), (\ref{eqv2 step2 const:c6}), (\ref{eq: c1-c3}), (\ref{eq: DC approxi})}.
	\end{aligned}
\end{equation}
 Problem (\ref{eq: rs step2 appro}) is a convex Quadratically Constrained Quadratic Program (QCQP), which can be solved using interior-point methods \cite{boyd2004convex}. The proposed SCA-based algorithm is summarized in Algorithm \ref{SCA algorithm}.
 \begin{algorithm}[h!]	
 	\textbf{Initialize}: $n\leftarrow0,t^{[n]}\leftarrow0$, $\mathbf{{P}}^{[n]},\theta^{[n]}, \bm{\alpha}^{[n]},\bm{\alpha}_c^{[n]},\bm{\rho}^{[n]},\bm{\rho}_c^{[n]}$\;
 	\Repeat{$|t^{[n]}-t^{[n-1]}|<\epsilon$}{
 		$n\leftarrow n+1$\;
 		Solve problem (\ref{eq: rs step2 appro}) using $\mathbf{{P}}^{[n-1]},\theta^{[n-1]}, \bm{\alpha}^{[n-1]},\bm{\alpha}_c^{[n-1]},\bm{\rho}^{[n-1]},\bm{\rho}_c^{[n-1]}$ and denote the optimal value of the objective function as $t^{\star}$ and the optimal solutions as $\mathbf{{P}}^{\star},\theta^{\star}, \bm{\alpha}^{\star},\bm{\alpha}_c^{\star},\bm{\rho}^{\star},\bm{\rho}_c^{\star}$\;
 		Update $t^{[n]}\leftarrow t^{\star}$, $\mathbf{P}^{[n]}\leftarrow \mathbf{P}^{\star}$,  $\theta^{[n]}\leftarrow\theta^{\star}$, $\bm{\alpha}^{[n]}\leftarrow\bm{\alpha}^{\star}$, $\bm{\alpha}_c^{[n]}\leftarrow\bm{\alpha}_c^{\star}$, $\bm{\rho}^{[n]}\leftarrow\bm{\rho}^{\star}$, $\bm{\rho}_c^{[n]}\leftarrow\bm{\rho}_c^{\star}$\;				
 	}	
 	\caption{Proposed SCA-based algorithm}
 	\label{SCA algorithm}		
 \end{algorithm}
%

\subsection{Convergence and Complexity}
\subsubsection{Convergence} The proposed SCA-based algorithm iteratively solves the approximated  problem (\ref{eq: rs step2 appro}) until convergence, where 
$\epsilon$ is the tolerance of convergence. 
\begin{prop} 
	For any feasible initial point, the proposed SCA-based algorithm is guaranteed to converge to a stationary point of problem (\ref{eq: rs step2}).  
\end{prop} 	

\textit{Proof}: SCA ensures monotonic improvement of $t$, i.e., $t^{[n]}\geq t^{[n-1]}$.	This is due to the fact that the solution generated by solving problem (\ref{eq: rs step2 appro}) at iteration $n-1$ is a feasible point of  problem (\ref{eq: rs step2 appro})  at iteration $n$. 
Due to the transmit power constraint (\ref{c4}), the sequence $\left\{t^{[n]}\right\}_{n=1}^{n=\infty}$ is bounded above, which implies that the convergence of the proposed SCA-based algorithm is guaranteed.  Next, we show that  the 
sequence of  $(\mathbf{{P}}^{[n]},\theta^{[n]}, \bm{\alpha}^{[n]},\bm{\alpha}_c^{[n]},\bm{\rho}^{[n]},\bm{\rho}_c^{[n]})$ converges to the set of stationary points of problem (\ref{eq: rs step2}). The proposed SCA-based algorithm is in fact an inner approximation algorithm in the nonconvex optimization literature \cite{marks1978general,KKTconvergence2013Li}.  This is proved by showing the equivalence of the KKT conditions of problem (\ref{eq: rs step2}) and problem (\ref{eq: rs step2 appro}) when the solution $(\mathbf{{P}},\theta, \bm{\alpha},\bm{\alpha}_c,\bm{\rho},\bm{\rho}_c)$ is equal to $(\mathbf{{P}}^{[n]},\theta^{[n]}, \bm{\alpha}^{[n]},\bm{\alpha}_c^{[n]},\bm{\rho}^{[n]},\bm{\rho}_c^{[n]})$.  Combining with the fact that the  approximations made in (\ref{eq: c1-c3}), (\ref{eq: DC approxi}) are asymptotically tight as $n\rightarrow \infty$ \cite{KKTconvergence2013Li}, we can obtain that the solution of the proposed SCA-based algorithm converges to the set of KKT points (which is also known as the stationary points) of problem (\ref{eq: rs step2}). 
\qed

\subsubsection{Complexity} At each iteration of the proposed SCA-based algorithm, the convex subproblem (\ref{eq: rs step2 appro}) is solved. Due to the exponential cone constraints (\ref{eqv2 step2 const:c5}) and (\ref{eqv2 step2 const:c6}),  problem (\ref{eq: rs step2 appro})  is a generalized nonlinear convex program. The nonlinear solvers such as  ``fmincon" in  the Matlab optimization toolbox can be adopted to solve the problem. An alternative efficient method is to approximate (\ref{eqv2 step2 const:c5}) and (\ref{eqv2 step2 const:c6})   by a sequence of Second Order Cone (SOC) constraints via the successive approximation method \cite{ben2001polyhedral,tervo2015optimalEE}. The resulting SOC Programming (SOCP) can be solved by using interior-point methods with computational complexity $\mathcal{O}([KN_t]^{3.5})$.  The total number of iterations required for the convergence is approximated as $\mathcal{O}(\log(\epsilon^{-1}) )$. Hence, the worst-case computational complexity is $\mathcal{O}(\log(\epsilon^{-1})[KN_t]^{3.5})$.

\textit{Remark 2: The Weighted Minimum Mean Square Error (WMMSE) algorithm with one dimensional exhaustive search proposed in \cite{Jian2019CRS} can also be adopted to solve problem (\ref{eq: rs step2}) for a given relaying user scheduling scheme, where $\theta$ is exhaustively searched over the range $(0, 1]$ and $\mathbf{P}(\theta), \mathbf{c}(\theta)$ are updated by using the WMMSE-based Alternative Optimization  (AO) algorithm for each $\theta$. Compared with the  algorithm proposed in \cite{Jian2019CRS}, our proposed SCA-based algorithm is more efficient as  $\theta, \mathbf{P}, \mathbf{c}$ are jointly optimized without one-dimensional exhaustive search. The computational complexity of the WMMSE and exhaustive search-based algorithm proposed in \cite{Jian2019CRS} is $\mathcal{O}(\delta^{-1}\log(\epsilon^{-1})[KN_t]^{3.5})$ where $\delta \in (0,1)$ is
the increment between two adjacent candidates of $\theta$. The set of candidate $\theta$ is assumed to be regularly-spaced over the range $(0, 1]$. $\epsilon$ is the convergence tolerance of the AO algorithm to update $\mathbf{P}(\theta), \mathbf{c}(\theta)$ for each $\theta$. It is assumed to be the same as in the proposed SCA-based algorithm. Therefore, the worst-case computational complexity of our proposed SCA-based algorithm is $\delta^{-1}$ folds lower than the algorithm proposed in \cite{Jian2019CRS}. 
}

\section{Numerical Results}
\label{sec: numerical results}
In this section, we evaluate the performance of the generalized $K$-user CRS using the proposed low-complexity algorithm. 

We compare the following relaying user  scheduling protocols for the design of $\mathcal{K}_1$ and  $\mathcal{K}_2$\footnote{ Under the assumption of perfect CSIT, the selection results of the proposed centralized and decentralized relaying protocols are the same for a given $K_1$. The influence of CSIT inaccuracy to the two different relaying protocols is worth to be studied as a future work. }:
\begin{itemize}
	\item \textbf{Optimal}---the optimal relaying protocol where the relaying user selection is performed centrally at BS by enumerating all possible relaying user combinations. The scheduling scheme with the highest max-min  rate is selected. It achieves the upper bound of the max-min  rate of all  relaying protocols but has the highest  selection complexity.
	\item \textbf{1 best relay}---the proposed relaying protocols when $K_1=1$. 
	\item \textbf{$\mathbf{K/2}$ best relays}---the proposed relaying protocols when $K_1=\frac{K}{2}$. If $K$ is  odd, $\frac{K}{2}$ is rounded up to the closest integer.
	\item \textbf{1 random relay}--- BS randomly selects one user  from $\mathcal{K}$ and informs the decision to all users via the ``RTS" packet. It has the lowest  selection complexity.
\end{itemize}

For a determined relaying user grouping $\mathcal{K}_1$ and  $\mathcal{K}_2$, we compare the following precoder, message split and time resource allocation algorithms:
 \begin{itemize}
 	\item \textbf{CRS: Proposed SCA}---the  cooperative RS model proposed in Section \ref{sec: system model} and the proposed SCA-based algorithm (Algorithm \ref{SCA algorithm}) is adopted to solve problem (\ref{eq: rs}). As discussed in Section \ref{sec: proposed algorithm}, its worst-case computational complexity is $\mathcal{O}(\log(\epsilon^{-1})[KN_t]^{3.5})$.
 	\item \textbf{CRS: WMMSE}---the  cooperative RS model proposed in Section \ref{sec: system model}, but the optimization problem (\ref{eq: rs}) is solved using the WMMSE algorithm  proposed in \cite{Jian2019CRS} with one-dimensional exhaustive search for $\theta$.  Its worst-case computational complexity  is $\mathcal{O}(\lambda^{-1}\log(\epsilon^{-1})[KN_t]^{3.5})$.
 	
 	\item \textbf{ERS: WMMSE}---Equal RS is a special instance of ``{CRS: WMMSE}" scheme when $\theta$ is fixed to 0.5. The time slot allocations for the direct and cooperative transmission phases are equal.  Its worst-case computational complexity is $\mathcal{O}(\log(\epsilon^{-1})[KN_t]^{3.5})$.
 		
 	\item \textbf{NRS: WMMSE}---Non-cooperative RS is also a  special instance of ``{CRS: WMMSE}" scheme when $\theta$ is fixed to 1. This is the RS scheme that has been investigated in \cite{RSintro16bruno,RS2016hamdi,RS2016joudeh,mao2017rate,mao2018EE} for MISO BC without cooperative transmission. The transmission completes at the end of the direct transmission phase and the relaying transmission is blocked.  Its worst-case computational complexity is $\mathcal{O}(\log(\epsilon^{-1})[KN_t]^{3.5})$.

 	\item \textbf{SDMA: WMMSE}--- the traditional multi-user linear precoding-based SDMA investigated in \cite{wmmse08}. There is no RS and no cooperative transmission (i.e., $\left\|\mathbf{p}_c\right\|^2=0$ and $\theta=1$).  Its worst-case computational complexity is $\mathcal{O}(\log(\epsilon^{-1})[KN_t]^{3.5})$.
 \end{itemize}

Different from \cite{Jian2019CRS} where only the specific channel realization is investigated for the two-user CRS, we consider the scenarios when each channel  $\mathbf{h}_k$  has independent and identically distributed (i.i.d.) complex Gaussian entries with a certain variance, i.e., $\mathcal{CN}(0, \sigma_k^2)$. The channel between user-$k$ and user-$j$ follows $h_{k,j}\sim\mathcal{CN}(0, \sigma_{k,j}^2)$. In the following simulation, we assume $\sigma_{k,j}^2=1,\forall k,j\in\mathcal{K},k\neq j$. The tolerance of all the algorithms are set to $\epsilon= 10^{-3}$. Without loss of generality, we assume that the transmit power at  BS and the relaying users in $\mathcal{K}_1$ are equal, i.e.,  $P_t= P_j, \forall j\in\mathcal{K}_1$. Following the literature \cite{Jian2019CRS, mao2017rate,RS2016hamdi},  the precoders of  the proposed SCA-based algorithm  are initialized by using  Maximum Ratio Transmission (MRT) combined with Singular Value Decomposition (SVD). The precoder $\mathbf{p}_k^{[0]}$ for the private stream $s_k$ is initialized as $\mathbf{p}_k^{[0]}=\sqrt{p_k}\frac{\mathbf{h}_k}{\left\Vert  \mathbf{h}_k\right\Vert}$, where $p_k=\frac{\beta P_t}{2}$ and $0 \leq \beta\leq 1$. The precoder $\mathbf{p}_c^{[0]}$ for the common message $s_c$ is initialized as $\mathbf{p}_c^{[0]}=\sqrt{p_c}\mathbf{u}_{c}$, where $p_{c}=(1-\beta)P_t$ and $\mathbf{u}_{c}$ is the largest left singular vector of the channel matrix $\mathbf{H}=[\mathbf{h}_1, \mathbf{h}_2,\ldots, \mathbf{h}_K]$. It is calculated by $\mathbf{u}_{c}=\mathbf{U}(:,1)$ where $\mathbf{H}=\mathbf{U}\mathbf{S}\mathbf{V}^{H}$. 
 $\{\alpha_{c,k}^{[0]},\alpha_{k}^{[0]},\rho_{c,k}^{[0]},\rho_{k}^{[0]}|k\in\mathcal{K}\}$ are initialized as $\rho_{c,k}^{[0]}=\frac{\left|\mathbf{h}_k^H\mathbf{p}_c^{[0]}\right|^2}{\sum_{j\in\mathcal{K}}\left|\mathbf{h}_k^H\mathbf{p}_j^{[0]}\right|^2+1}$, $\rho_{k}^{[0]}=\frac{\left|\mathbf{h}_k^H\mathbf{p}_k^{[0]}\right|^2}{\sum_{j\in\mathcal{K},j\neq k}\left|\mathbf{h}_k^H\mathbf{p}_j^{[0]}\right|^2+1}$, $\alpha_{c,k}^{[0]}=\log_2(1+\rho_{c,k}^{[0]})$, $\alpha_{k}^{[0]}=\log_2(1+\rho_{k}^{[0]})$,  respectively.  $\theta^{[0]}$ is initialized as 0.8. The precoder initialization of the WMMSE algorithm is the same as the proposed SCA-based algorithm.  $\theta$ is searched with increment $\delta=0.1$ in the ``CRS: WMMSE" algorithm. Therefore, the precoders and message split are optimized by using the WMMSE algorithm 10 times for  each value of $\theta$  selected from the set $[0.1,0.2,\ldots,1]$. The MATLAB toolbox YALMIP \cite{yalmip} is used to solve the QCQP problem (\ref{eq: rs step2 appro}).
 
\begin{figure}
	\centering
	\begin{subfigure}{0.45\textwidth}
		\centering
		\includegraphics[width=3.3in]{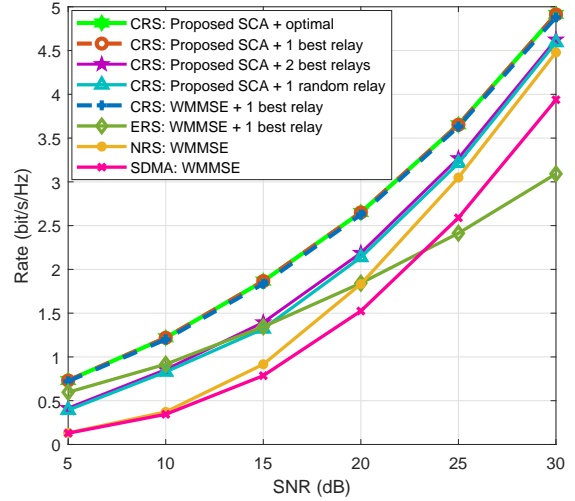}
		\centering
		\caption{Underloaded case ($N_t=4$),  $\sigma_1^2=1$, $\sigma_2^2=0.3$, $\sigma_3^2=0.1$}
		\label{fig: snrvsrateNt4bias1103}
	\end{subfigure}%
	\\
	\begin{subfigure}{0.45\textwidth}
		\centering
		\includegraphics[width=3.3in]{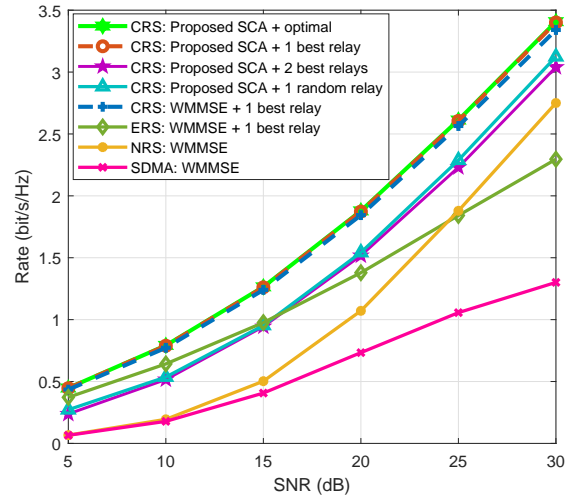}
		\caption{Overloaded case ($N_t=2$), $\sigma_1^2=1$, $\sigma_2^2=0.3$, $\sigma_3^2=0.1$ }
		\label{fig: snrvsrateNt2bias10301}
	\end{subfigure}%
	\\
	\begin{subfigure}{0.45\textwidth}
		\centering
		\includegraphics[width=3.3in]{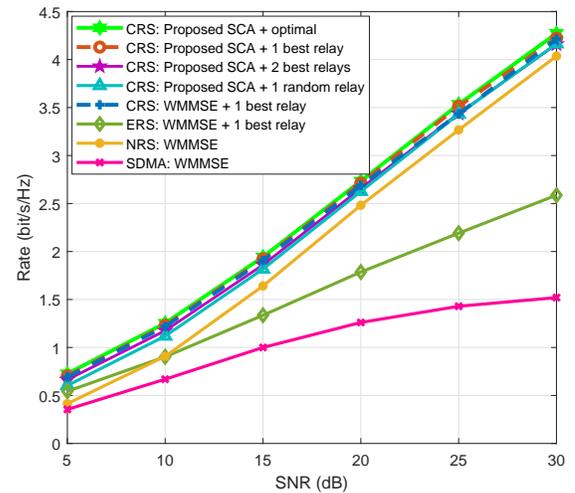}
		\centering
			\caption{Overloaded case ($N_t=2$), $\sigma_1^2=1$, $\sigma_2^2=1$, $\sigma_3^2=0.3$}
		\label{fig: snrvsrateNt2bias1103}
	\end{subfigure}%
	\caption{The rate performance versus SNR comparison of different
		strategies, averaged over 100 random channel realizations. $K=3$.}
	\label{fig: snrvsrate}
\end{figure}

Fig. \ref{fig: snrvsrate} shows the average max-min rates of different strategies used in two stages versus SNR at BS over 100 random channel realizations with varied transmit antennas and channel strength disparities among users ($K=3$). The variances of user channels are set to  $\sigma_1^2=1$, $\sigma_2^2=0.3$, $\sigma_3^2=0.1$ in subfigure (a) and (b) while $\sigma_1^2=1$, $\sigma_2^2=1$, $\sigma_3^2=0.3$ in subfigure (c). The number of transmit antennas is 4 in subfigure (a)  (which is an underloaded case) while there is 2 transmit antennas in subfigure (b) and (c) (which are overloaded cases). 

Comparing the relaying protocols for the given joint optimization algorithm  ``CRS: Proposed SCA" within each subfigure of Fig. \ref{fig: snrvsrate}, we observe that  ``1 best relay" achieves  almost the same rate performance as  the optimal relaying scheduling algorithm. In comparison, ``${K/2}$ best relays", which is  equal to ``2 best relays" when $K=3$,  deteriorates the max-min rate and is  closer to the performance of the baseline "1 random relay". The users served as relaying users in $\mathcal{K}_1$ cannot achieve any common rate improvement while their private rates deteriorate due to the time slot allocation. To enhance the rate among users, reducing the size of $\mathcal{K}_1$ is preferred as more users can benefit from the cooperative transmission. Under the given ``1 best relay" scheduling protocol, we compare the  algorithms of optimizing precoders, message split and time  resource allocation. In all subfigures, the proposed ``CRS: Proposed SCA" algorithm achieves the same or  even higher rate performance when compared with ``CRS: WMMSE", but at a  much lower  computational complexity. In Fig. \ref{fig: snrvsrate}(a), the  achievable max-min rate of ``CRS: Proposed SCA"  averaged across all SNRs  attains  relative rate gain  of 143.4\%, 41.4\%  and 166.1\%  when compared with  that of ``NRS: WMMSE", ``ERS: WMMSE" and ``SDMA: WMMSE", respectively.  Therefore, the proposed CRS with SCA-based algorithm achieves  explicit rate improvement over all the existing transmission schemes. 

We further compare the rates achieved by different strategies in Fig. \ref{fig: snrvsrate}(a) and Fig. \ref{fig: snrvsrate}(b) with four or two transmit antennas. The  relative rate gain of  ``CRS: Proposed SCA" (with ``1 best relay" scheduling protocol) over ``NRS: WMMSE" averaged across all SNRs increases from  143.4\%  in Fig. \ref{fig: snrvsrate}(a) to  192.9\%  in Fig. \ref{fig: snrvsrate}(b) as the number of transmit antennas decreases. This is due to the fact that  multi-user interference received at each user in the overloaded case is more severe than that in the underloaded case. A larger portion of users' private streams  is encoded in the common stream for each user to decode. As the relaying users in CRS re-transmit the  common stream, the amount of interference to be decoded at each user is further enhanced. Hence, CRS has a greater capability to manage the interference and it is more suited to the scenarios when users suffer from stronger multi-user interference. We also observe that the rate gap between the  proposed ``CRS: Proposed SCA" (with ``1 best relay") and ``SDMA: WMMSE" increases as the number of transmit antennas decreases. The relative rate gain of ``CRS: Proposed SCA"  over ``SDMA: WMMSE" averaged over SNRs increases from  166.1\% in Fig. \ref{fig: snrvsrate}(a)  to   266.1\% in Fig. \ref{fig: snrvsrate}(b).   SDMA is only suited to the underloaded scenario.  As $N_t$ decreases, SDMA cannot manage the multi-user interference coming from all  other users  and the rate saturates at high SNR. In comparison, RS assisted transmission strategies are more robust to the network load thanks to its ability to partially decode the interference and partially treat interference as noise. This observation coincides with the observations in \cite{RS2016hamdi,mao2017rate}. We should also notice that, with the assistance of cooperative transmission, the gain achieved by CRS over SDMA is larger than that achieved by NRS over SDMA at low SNR. When SNR is 10 dB, the relative rate gains between CRS and SDMA in three subfigures are 253.8\%, 342.9\%, 83.8\%, respectively while the relative rate gains between NRS and SDMA in three subfigures are 8.04\%, 9.2\%, 35.6\%, respectively.

Comparing the rate achieved by the strategies under the same network load but different channel strength disparities among users (subfigure (b) and (c) in Fig. \ref{fig: snrvsrate}), we observe that the rate region gap between ``CRS: Proposed SCA" (with ``1 best relay")  and ``NRS: WMMSE" increases from an average of  23.9\%  in Fig. \ref{fig: snrvsrate}(c) to an average of  192.9\%  in Fig. \ref{fig: snrvsrate}(b) as the channel strength  disparities among users increases.  Without cooperative transmission, the achievable common rate in NRS  is limited by the common rate of the worst-case user. The loss of common rate becomes more severe when there is a larger channel strength disparity.  In comparison, CRS further enhances the common rate of the worst-case user by retransmitting the common stream to that user.  When the user channel strength disparities among users are  small, $\theta$ is much closer to 1. Therefore, the benefits of using CRS over NRS fall off in such circumstance. 

\begin{figure}[t!]
	\centering
	\includegraphics[width=3.3in]{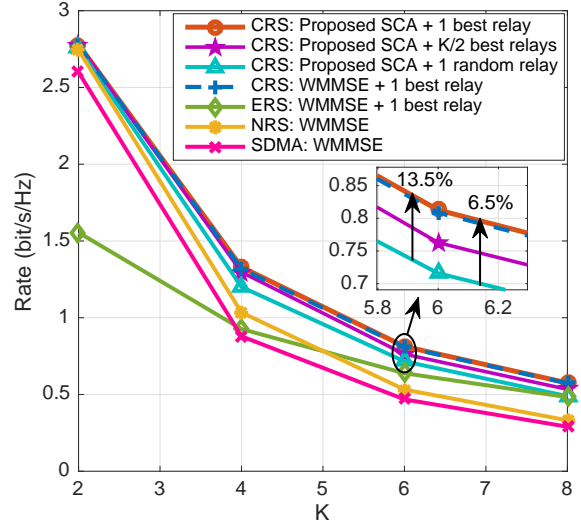}
	\caption{The rate performance versus the number of users $K$  of different strategies, averaged over 100 random channel realizations. SNR=10 dB.}
	\label{fig: ratevsNuser}
\end{figure}

Fig. \ref{fig: ratevsNuser} shows the average rate of different strategies versus number of users $K$ over 100 random channel realizations when SNR is 10 dB. The variances of the user channels  uniformly decrease from 1 with stepsize $\frac{1}{K}$. For example, $\sigma_1^2=1, \sigma_2^2=0.5$ when $K=2$ and $\sigma_1^2=1, \sigma_2^2=0.75, \sigma_3^2=0.5, \sigma_4^2=0.25$  when $K=4$. As the number of users increases, the optimal relaying protocol becomes computational prohibitive since $\sum_{k=1}^{K-1}\binom{K}{k}$ possible scheduling groups need to be considered. Under the given joint optimization algorithm ``CRS: Proposed SCA", ``1 best relay" is able to achieve an average of 57.83\% rate improvement over ``$K/2$ best relays" as well as  an average of 13.5\% rate improvement over ``1 random relays". From the observations of the scheduling methods in Fig. \ref{fig: snrvsrate} and Fig. \ref{fig: ratevsNuser}, we conclude that a single relaying user selection based on channel strength is effective enough  in the $K$-user CRS-assisted transmission network. 

As the number of users increases, the rate loss of NRS  becomes more obvious while the performance of ERS improves when compared with CRS. It implies that the optimal $\theta$ decreases as the number of users increases. This is due to the fact that multi-user interference increases with the number of users, and a larger portion of the user messages is transmitted via the common stream. CRS benefits from the ability to enhance the common rate and therefore boosts the max-min fairness of the system. We can also observe that CRS with one random relaying user selection achieves non-negligible rate improvement over NRS and ERS.   Comparing the RS-based schemes with the baseline SDMA, we confirm the superiority of RS over traditional SDMA in both underloaded ($K\leq4$ in Fig. \ref{fig: snrvsrate}) and overloaded ($K>4$ in Fig. \ref{fig: snrvsrate})  MISO BC as demonstrated in the literature \cite{RS2016hamdi,RS2016joudeh,hamdi2017bruno,mao2017rate}\footnote{We should also notice that CRS has been shown to outperform C-NOMA where the message to be decoded first by all users is re-transmitted from the relaying users to other users. Similar observations are obtained in the non-cooperative transmission scenarios. Readers are referred to  \cite{Jian2019CRS} for more information about the rate gain of  CRS over C-NOMA as well as \cite{mao2017rate,mao2019swipt,mao2019TCOM,bruno2019wcl} for  the  rate improvement  of NRS over conventional NOMA without cooperative transmission. }.

\begin{figure}[t!]
	\centering
		\begin{subfigure}{0.48\textwidth}
				\centering
		\includegraphics[width=3.3in]{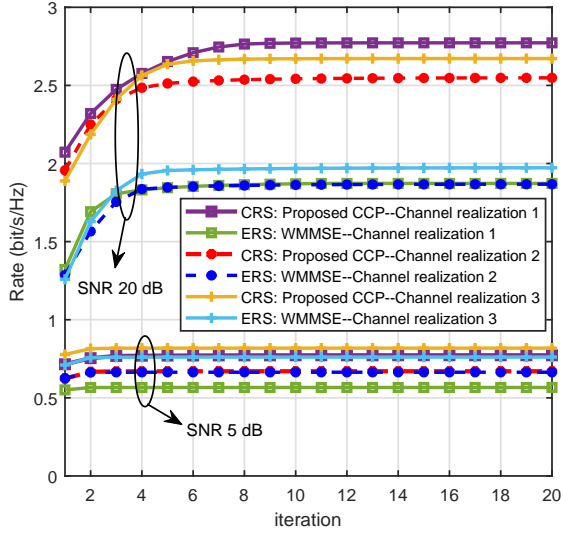}
		\caption{Three different channel realizations with initialization method based on MRT-SVD.}
		\label{fig: convergence_diffChannel}
	\end{subfigure}%
~\\
	\begin{subfigure}{0.48\textwidth}
				\centering
		\includegraphics[width=3.3in]{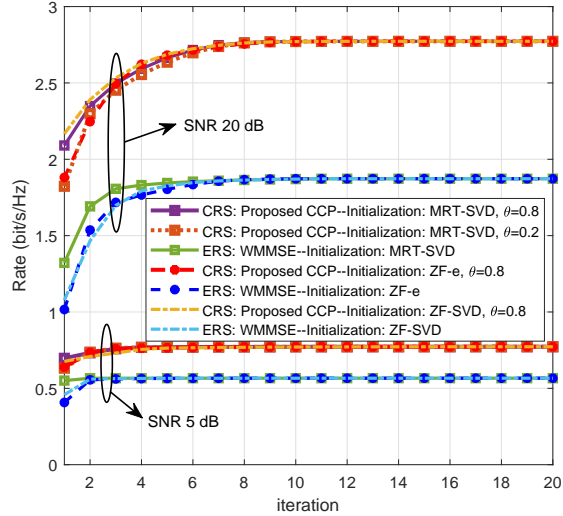}
		\centering
		\caption{Three different initializations of $\mathbf{P}$ (namely, MRT-SVD, ZF-$e$, ZF-SVD) and two different initializations of $\theta$ ($\theta=0.2,0.8$) for one random channel realization.}
	\end{subfigure}%
	\caption{	Convergence comparison of  different algorithms for different  random channel realizations and different initializations of $\mathbf{P}$ and $\theta$. $K=3$.}
	\label{fig: convergence}
\end{figure}

As ``CRS: Proposed SCA", ``NRS: WMMSE" and ``ERS: WMMSE" maintain the same low computation complexity, we further compare the number of iterations required for the proposed SCA-based algorithm and the WMMSE algorithm to converge.  Fig. \ref{fig: convergence} illustrates the convergence results of ``CRS: Proposed SCA" and ``ERS: WMMSE" algorithms for different random channel realizations as well as different initializations of  $\mathbf{P}$ and $\theta$ when SNR is equal to 5 dB or 20 dB.  ``1 best relay" relaying protocol is adopted for both CRS and ERS. In Fig. \ref{fig: convergence}(a), the rate achieved by the two algorithms are compared for three different random channel realizations.  The initialization method is inline with the method used for Fig. \ref{fig: snrvsrate} and Fig.  \ref{fig: ratevsNuser}, where the precoders $\mathbf{p}_c$ and $\mathbf{p}_k, k\in\mathcal{K}$ are respectively initialized by MRT and SVD while  the time slot allocation $\theta$ is initialized as 0.8. 
It is evident that both algorithms converge within a few iterations. Though the proposed SCA-based algorithm jointly optimizes $\theta,\mathbf{P},\mathbf{c}$, it is able to converge quickly within a few iterations at both high and low SNR. The convergence speed is similar to WMMSE which only jointly optimizes $\mathbf{P},\mathbf{c}$. 
 In Fig. \ref{fig: convergence}(b), the rate achieved by the two algorithms with different initialization of  $\mathbf{P}$ and $\theta$ are compared for one certain random channel realization. Three different initialization methods of $\mathbf{P}$ are used, namely, MRT-SVD, ZF-$e$ and ZF-SVD.  ZF-$e$ is a DoF-motivated precoder design as adopted in \cite{RS2016hamdi} where $\mathbf{p}_k^{[0]}=\sqrt{p_k}\bar{\mathbf{p}}_k^{[0]}, \forall k\in\mathcal{K}$ and $\mathbf{p}_c^{[0]}=\sqrt{p_c}\bar{\mathbf{p}}_c^{[0]}$. $[\bar{\mathbf{p}}_1^{[0]},\ldots,\bar{\mathbf{p}}_K^{[0]}]$ are normalized ZF beamforming vectors constructed by using $\mathbf{H}=[\mathbf{h}_1, \ldots,\mathbf{h}_K]$ and $\bar{\mathbf{p}}_c^{[0]}$ is a randomly generated normalized vector. ZF-SVD is a modification of ZF-$e$ where $\mathbf{p}_c^{[0]}$ is initialized by SVD. $\theta^{[0]}$ is initialized as 0.2 or 0.8. Though the initializations of $\mathbf{P}$ and $\theta$ influence the convergence speed,  both algorithms still converge within limited iterations and different initialization methods  have very close convergence rate at both high and low SNR. 
 
 \begin{figure}[t!]
 	\centering
 	\includegraphics[width=3.3in]{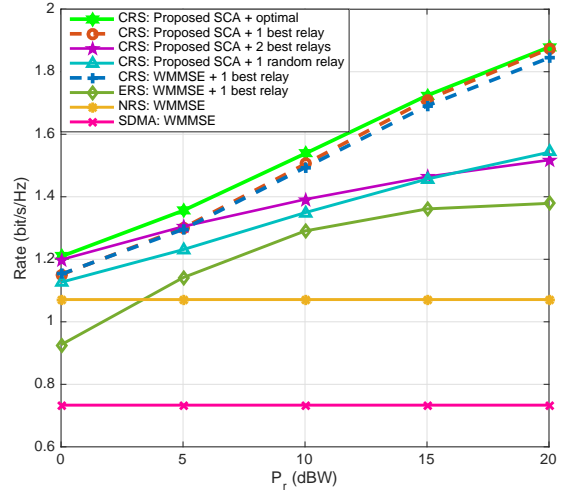}
 	\caption{The rate performance versus the transmit power constraint at the relaying users  of different strategies, averaged over 100 random channel realizations. $P_t=20$  dBW, $N_t=2$, $K=3$, $\sigma_1^2=1$, $\sigma_2^2=0.3$, $\sigma_3^2=0.1$.}
 	\label{fig: ratevsPr}
 \end{figure}
 
  The  results shown in Fig. \ref{fig: snrvsrate}--Fig. \ref{fig: convergence} assume equal  transmit power constraints at BS and the relaying users, i.e.,  $P_t= P_j, \forall j\in\mathcal{K}_1$.  We further study the impact of  different transmit power constraints at the relaying users on the rate performance in Fig. \ref{fig: ratevsPr}. The transmit power constraint at BS is fixed to $P_t=20$ dBW while the transmit power constraints at the relaying users are assumed to be equal to $P_r=P_j, \forall j\in\mathcal{K}_1$ and $P_r$ varies from 0 dBW to 20 dBW. As is illustrated in Fig. \ref{fig: ratevsPr}, the non-cooperative SDMA and NRS strategies remain constant as $P_r$ increases since  both schemes only consider the use of  the direct transmission phase from BS to the users. $P_r$ is not utilized at all. The cooperative transmission achieves a higher rate performance at the sacrifice of investing additional transmit power at the relaying users. We observe that as $P_r$ increases, the rate performance of the proposed ``1 best relay" scheduling protocol is much closer to that of the optimal one.  It  achieves    95.6\%  of the average rate of the optimal scheduling protocol and 105.6\% relative rate improvement over SDMA. Therefore, it is the preferred selection protocol when considering  the trade-off between the total power consumption ($P_t+K_1P_r$) and max-min rate.

\section{Conclusions}
\label{sec: conclusion}
To conclude, we investigate the  max-min fairness of $K$-user CRS for MISO BC with user relaying by designing the precoder, the RS message split and time slot allocation as well as the relaying user selection scheme in order to maximize the minimum QoS rate among users subject to a total transmit power constraint. We propose a two-stage low-complexity algorithm to solve the problem.  In the first stage, channel-gain based centralized and decentralized relaying protocols are proposed and compared. In the second stage, a  joint time slot allocation, precoder and message split optimization algorithm based on SCA method is proposed to solve the max-min fairness problem. Numerical results show that the proposed relaying protocols with single relaying user selection achieves  nearly the same rate performance  as the optimal relaying scheduling algorithm. RS simplifies the scheduling procedure in the cooperative transmission networks and  only the CSI between BS and the user is required at each user for decentralized relaying user scheduling. We further show through numerical results  that the proposed SCA-based algorithm solves the problem much more efficiently than the algorithm in the literature, and without any rate loss. Finally,  the influence of SNR, the number of transmit antennas, channel strength disparities among users as well as the number of  users to the CRS performance is studied.  We find that as the number of  transmit antenna decreases,  or the channel strength disparity among users increases or the number of serving users increases or at  low SNR, the rate improvement of CRS over  non-cooperative RS and SDMA become more significant. Due to its ability to enhance the transmission rate of the common stream, CRS is more capable of managing strong multi-user interference  and disparity of channel strengths. Hence, we draw the conclusion that $K$-user CRS is  superior than existing transmission schemes and has a great potential to  improve the system performance  of future communication networks. 

\section{Acknowledgement}
\label{sec: acknowledge}
This project was partially funded by the National Plan for Science, Technology and Innovation (MAARIFAH), King Abdulaziz City for Science and Technology, Kingdom of Saudi Arabia, Award Number (11-INF1951-02), and by the
U.K. Engineering and Physical Sciences Research Council (EPSRC) under grant EP/R511547/1.

\bibliographystyle{IEEEtran}
\bibliography{reference}	

\end{document}